\documentclass{amsart}
\newtheorem{theorem}{Theorem}[section]
\newtheorem{corollary}[theorem]{Corollary}
\newtheorem{proposition}[theorem]{Proposition}
\newtheorem{lemma}[theorem]{Lemma}
\newtheorem{definition}[theorem]{Definition}

\usepackage{amsmath}
\usepackage{epic}
\usepackage{eepic}
\usepackage{epsfig}
\usepackage{color}
\def\x{\fil{y}x}
\def\y{\fil{x}y}
\def\ys{y\fil{x}}
\def\xs{\fil{y\st}x}
\def\yss{\fil{z\st}y\fil{x}}
\def\z{z\fil{y\fil{x}}}
\def\dvrstep{\mathtt{dvr}}

\title[Parallel/Distributed Optimal Reduction]{
PELCR: Parallel Environment for Optimal Lambda-Calculus Reduction} 
\author[M. Pedicini, F. Quaglia]{Marco Pedicini\\
Istituto per le Applicazioni del Calcolo ``M.~Picone'' - CNR\\
\and \\
Francesco Quaglia\\
Universit\`a di Roma ``La Sapienza''
}

\def\EX{\hbox{\textsf{EX}}}

\def\st{{}^\star}
\def\pl{^+}
\def\Ls{\hbox{$\mathsf{L}\st$}}
\def\KLs{\hbox{$\Z[\Ls]$}}
\def\fil#1{[#1]}
\def\fin#1{\langle#1\rangle}
\def\cl#1#2#3#4#5{\mbox{\bf #1}^{#2}(#3;#4;#5)}
\def\al{\alpha}\def\ga{\gamma}\def\ba{\beta}\def\om{\omega}
\def\W{\hbox{W}}

\def\Z{\mathbf{Z}}

\begin{document}
\bibliographystyle{alpha} 

\begin{abstract}
In this article we present the implementation of an environment
supporting L\'evy's \emph{optimal reduction} for the
$\lambda$-calculus \cite{Lev78} on parallel (or distributed) computing
systems.  In a similar approach to Lamping's one in \cite{Lamping90},
we base our work on a graph reduction technique known as
\emph{directed virtual reduction} \cite{DPR97} which is actually a
restriction of Danos-Regnier virtual reduction \cite{DanosRegnier93}.
  
The environment, which we refer to as PELCR (Parallel Environment for
optimal Lambda-Calculus Reduction) relies on a strategy for directed
virtual reduction, namely {\em half combustion}.
While developing PELCR we have adopted both a message
aggregation technique, allowing a reduction of the communication
overhead, and a fair policy for distributing dynamically originated
load among processors.  Additionally, we have used a set of other
optimizations, e.g. allowing the maintenance of relatively low size
for the manipulated data structures so not to incur problems related
to their management at the application level or due to the management
of large process memory images at the operating system level.

We also present an experimental study demonstrating the ability of
PELCR to definitely exploit parallelism intrinsic to $\lambda$-terms
while performing the reduction. 
We show how PELCR allows achieving up to 70/80\% of the ideal speedup
on last generation multiprocessor computing systems.  As a last note,
the software modules have been developed with the {\tt C} language and
using a standard interface for message passing, i.e. MPI, thus making
PELCR itself a highly portable software package.
\end{abstract}



\maketitle

\footnotetext{An earlier version of this article by the same authors, 
with title ``A Parallel Implementation for Optimal Lambda-Calculus Reduction'' 
appeared in {\em 
Proc. of the 2nd ACM SIGPLAN Int. Conference on 
Principles and Practice of Declarative Programming (PPDP 2000)
}.

Authors' addresses: M. Pedicini, Istituto per le Applicazioni del Calcolo ``M.~Picone'',
Consiglio Nazionale delle Ricerche,
Viale del Policlinico 137, 00161 Roma, Italy, email
{\tt marco@iac.cnr.it};
F. Quaglia, Dipartimento di Informatica e
Sistemistica, Universit\`a di Roma ``La Sapienza'', Via Salaria 113,
00198 Roma, Italy, email {\tt quaglia@dis.uniroma1.it}.
}

\maketitle

\section{Introduction}

Jean-Jacques L\'evy formally characterized the meaning of the word
\emph{optimal} relatively to a reduction strategy for
$\lambda$-calculus, referring to it as the property that the strategy
reaches the normal form (if it exists) and does not duplicate the work
of reducing similar $\beta$-redexes \cite{Lev78}.

This characterization was formalized in terms of \emph{families of
redexes} that is, redexes with the same origin, possibly this origin
being a virtual one in the sense that two families coming in a
configuration producing a new redex originate a new family.
Redexes belonging to different families cannot be successfully shared
during reduction; whereas for two redexes in the same family, one could
find an optimal strategy (i.e. reducing all of them in a single step).

Data structures suitable for an implementation of optimal reduction
were presented a long time later \cite{Lamping90}; the outcome
reduction technique introduced by J.~Lamping, known as \emph{sharing
reduction}, relies on a set of graph rewriting rules.

In \cite{AbadiGonthierLevy92}, Lamping's sharing reduction was
proved to be a way to compute Girard's execution formula, which is an
invariant of closed functional evaluation obtained from the ``Geometry
of Interaction'' interpretation of $\lambda$-calculus
\cite{Girard89a}.  This result stirred the research in the
field of optimal reduction. Specifically, in \cite{DanosRegnier93} a
graphical local calculus, namely {\em virtual reduction} (VR), was
defined as a mechanism to perform optimal reduction by computing the
Girard's execution formula. Such a calculus was later refined in
\cite{DPR97}, by the introduction of a new graph rewriting technique
 known as {\em directed virtual reduction} (DVR). The authors also defined
 a strategy to perform DVR, namely {\em combustion}, which
 simplifies the calculus and can simulate individual steps of sharing
 reduction.

In this article we describe a technique for the implementation of
functional calculi.  This technique exploits both \emph{locality and
asynchrony} of the computation which is typical in interaction nets
(\cite{Lafont90}) and derives from the fine decomposition of the
$\lambda$-calculus $\beta$-rule obtained through the analysis provided
by the Geometry of Interaction. Specifically, we present the
implementation of a Parallel Environment for optimal Lambda-Calculus
Reduction (PELCR) which relies on DVR and on a new strategy to perform
DVR that will be referred to as
\emph{half-combustion} (HC).

Let us stress that any interpreter of an ML-like functional language
based on our technique ensures the execution of programs in a parallel (or
distributed) environment in a way completely transparent to the user.

To the best of our knowledge, our work is the first attempt for
parallel implementations of optimal $\lambda$-calculus reduction.
Actually in \cite{Mackie97} issues on the possibility of parallel
implementations for Lafont's interaction nets are discussed. In that
work Mackie is faced to problems of load balancing and fine grain
parallelism. The solution proposed in \cite{Mackie97} is a
\emph{static analysis} of the initial interaction net which aims at
setting up a favorable initial distribution of the nodes among
processors.  His work is related to optimal reduction since optimal
rules (e.g. in \cite{AbadiGonthierLevy92}) define an interaction
system \cite{Lafont90}. However, contrarily to our work, it does not
focus on optimal reduction.  Another fundamental difference between
our work and Mackie's study is that our approach is dynamic: load
distribution is decided at run-time and the message passing overhead
is controlled dynamically as well. Also, our implementation embeds a
set of other optimizations further allowing improved run-time
behavior, e.g. for what concerns memory performance at both the
application level and the operating system level.

The implementation has been developed with the {\tt C} language using
a standard interface, namely MPI, for supporting message passing
functionalities among processes involved in the computation. These
peculiarities make PELCR a highly portable software package, easy to
install on a wide set of, possibly heterogeneous, computing platforms.
 
We also report the results of an experimental evaluation of our
software package. By the experimental data, we show how it has the
ability to definitely exploit any form of parallelism intrinsic to the
reduction of $\lambda$-terms. As a result, we obtain up to 70/80\% of
the ideal speedup, i.e. the ideal acceleration as compared to a
sequential case, while performing $\lambda$-term reductions on last
generation multiprocessor systems. This, in its turn, also allows
decreasing the wall-clock time for the reduction from several tens of
seconds to few seconds. This points out how PELCR has the potential to
cope with response time requirements for the satisfaction of an
interactive end-user even in case of jobs that would require large
computation time if executed in a classical sequential fashion.

We analyze the problem (parallel implementation of functional calculi)
from a pragmatic point of view, and the theory of (directed) virtual
reduction is here considered mainly for how it can give rise to
parallel dynamics.  However, while recalling such a theory, we also
propose a few optimization rules allowing an increase in the
effectiveness of the DVR approach (see Section~\ref{optrules}), which
have been taken into account while developing the implementation. The
remainder of the article is structured as follows. In
Section~\ref{background} we recall DVR.  In Section~\ref{hc} the HC
strategy for DVR is introduced. In Section~\ref{algorithm} we report
the description of our implementation.  The experimental results are
reported in Section~\ref{performance}.

\emph{Acknowledgments.} The project of a parallel and optimal
interpreter for $\lambda$-calculus started as a joint effort between
the University of Paris 7 and the ``Istituto per le Applicazioni del
Calcolo'' in Rome (see ``Optimal and parallel evaluations in
functional languages" CNR/CNRS - Bilateral Project n.3132 - 1996/97);
some of the ideas used in our implementation arose thanks to
discussions of the first author with Vincent Danos. The authors also
wish to thank Carlo Giuffrida for his support while developing some
software modules.

\section{From Lambda-Terms to Directed Virtual Reduction}
\label{background} 

As pointed out in the introduction we deal with an evaluator for
$\lambda$-terms based on DVR, to be executed on parallel/distributed
computing system. The pioneering ideas contained in L\'evy's work on
optimal reduction where finally realized by Lamping and then related
to semantical questions about operational aspects of computations. In
fact almost at the same time Girard gave the foundations of an
outstanding mathematical base for the study of operational semantics.

Just to enumerate them we should cite Lamping's first work on sharing
reduction, the connection with Geometry of Interaction discovered by
Gonthier and finally the work of Danos-Regnier on VR and DVR.  There
is no way to get a complete and self-contained presentation of all
this material, therefore, for a complete survey about the optimal
implementation of functional programming languages we refer the reader
to \cite{AspertiGuerrini98}.

Here we shortly recall VR and the Geometry of Interaction
(Section~\ref{goi}); then we will give a full presentation of DVR
(Section~\ref{pippo}), with the introduction of some properties
(Section~\ref{optrules}) which will be taken into account in the
implementation. To ease the comprehension of this reduction technique
and to make a more direct connection with Lamping's graphs we finally
present an encoding of such graphs into directed virtual nets
(Section~\ref{trans}).

\begin{figure}
\centerline{
\begin{picture}(0,0)%
\includegraphics{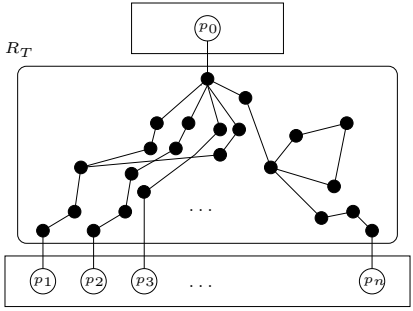}%
\end{picture}%
\setlength{\unitlength}{2092sp}%
%
\gdef\SetFigFont#1#2#3#4#5{%
	\fontsize{#1}{#2pt}%
  \fontfamily{#3}\fontseries{#4}\fontshape{#5}%
  \selectfont}%
\begin{picture}(4824,3624)(589,-3373)
\put(601,-361){\makebox(0,0)[lb]{\smash{{\SetFigFont{6}{7.2}{\rmdefault}{\mddefault}{\updefault}{\color[rgb]{0,0,0}$R_T$}%
}}}}
\put(2776,-3136){\makebox(0,0)[lb]{\smash{{\SetFigFont{6}{7.2}{\rmdefault}{\mddefault}{\updefault}{\color[rgb]{0,0,0}$\dots$}%
}}}}
\put(2776,-2236){\makebox(0,0)[lb]{\smash{{\SetFigFont{6}{7.2}{\rmdefault}{\mddefault}{\updefault}{\color[rgb]{0,0,0}$\dots$}%
}}}}
\put(1549,-3080){\makebox(0,0)[lb]{\smash{{\SetFigFont{5}{6.0}{\rmdefault}{\mddefault}{\updefault}{\color[rgb]{0,0,0}$p_2$}%
}}}}
\put(2149,-3080){\makebox(0,0)[lb]{\smash{{\SetFigFont{5}{6.0}{\rmdefault}{\mddefault}{\updefault}{\color[rgb]{0,0,0}$p_3$}%
}}}}
\put(940,-3080){\makebox(0,0)[lb]{\smash{{\SetFigFont{5}{6.0}{\rmdefault}{\mddefault}{\updefault}{\color[rgb]{0,0,0}$p_1$}%
}}}}
\put(4851,-3083){\makebox(0,0)[lb]{\smash{{\SetFigFont{5}{6.0}{\rmdefault}{\mddefault}{\updefault}{\color[rgb]{0,0,0}$p_n$}%
}}}}
\put(2896,-77){\makebox(0,0)[lb]{\smash{{\SetFigFont{5}{6.0}{\rmdefault}{\mddefault}{\updefault}{\color[rgb]{0,0,0}$p_0$}%
}}}}
\end{picture}%
}
\caption{An explicative picture for execution paths.}\label{b2b}
\end{figure}

The basic ingredient in Gonthier and Danos-Regnier works is the use of
the invariance of the execution formula as a consistency criterion for
the reduction technique. The execution formula associated with a term
$T$ with free variables $\{x_1 , \dots, x_n\}$ is given by the set of
its border-to-border weighted straight paths in its dynamic graph
$R_T$. Any node $p_i$ in the border node 
is either
associated with one free variable $x_i$ or, if $i=0$, $p_0$ represents
the root of the term, see for example Figure \ref{b2b}.  A straight
path in a directed graph is a path that never bounces back in the same
edge.

The execution formula of $R_T$ is: $$\EX(R_T) = \sum_{\phi_{ij} \in
{\mathcal P}(R_T)} W(\phi_{ij})$$ where $W(.)$ is a morphism from the
involutive category of paths ${\mathcal P}(R_T)$ to the monoid of the
Geometry of Interaction, so that for any straight path $\phi_{ij}$
from $p_i$ to $p_j$, $W(\phi_{ij})$ is an element of that monoid.

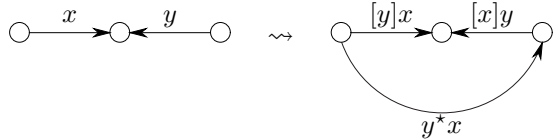
\begin{figure}[ht]
  \begin{center}
    \leavevmode
\setlength{\unitlength}{0.00070333in}
\begingroup\makeatletter\ifx\SetFigFont\undefined%
\gdef\SetFigFont#1#2#3#4#5{%
  \reset@font\fontsize{#1}{#2pt}%
  \fontfamily{#3}\fontseries{#4}\fontshape{#5}%
  \selectfont}%
\fi\endgroup%
{\renewcommand{\dashlinestretch}{30}
\begin{picture}(4066,1035)(0,-10)
\put(3234.500,985.071){\arc{1592.145}{0.3475}{2.7941}}
\blacken\path(3970.346,590.956)(3983.000,714.000)(3913.932,611.386)(3954.397,635.020)(3970.346,590.956)
\put(83,789){\ellipse{150}{150}}
\put(3233,789){\ellipse{150}{150}}
\put(3983,789){\ellipse{150}{150}}
\put(2483,789){\ellipse{150}{150}}
\put(1583,789){\ellipse{150}{150}}
\put(833,789){\ellipse{150}{150}}
\path(1508,789)(908,789)
\blacken\path(1028.000,819.000)(908.000,789.000)(1028.000,759.000)(992.000,789.000)(1028.000,819.000)
\path(2558,789)(3158,789)
\blacken\path(3038.000,759.000)(3158.000,789.000)(3038.000,819.000)(3074.000,789.000)(3038.000,759.000)
\path(3908,789)(3308,789)
\blacken\path(3428.000,819.000)(3308.000,789.000)(3428.000,759.000)(3392.000,789.000)(3428.000,819.000)
\path(158,789)(758,789)
\blacken\path(638.000,759.000)(758.000,789.000)(638.000,819.000)(674.000,789.000)(638.000,759.000)
\put(2033,714){\makebox(0,0)[b]{\smash{{{\SetFigFont{10}{14.4}{\rmdefault}{\mddefault}{\updefault}$\leadsto$}}}}}
\put(458,864){\makebox(0,0)[b]{\smash{{{\SetFigFont{10}{14.4}{\rmdefault}{\mddefault}{\updefault}$x$}}}}}
\put(1208,864){\makebox(0,0)[b]{\smash{{{\SetFigFont{10}{14.4}{\rmdefault}{\mddefault}{\updefault}$y$}}}}}
\put(2858,864){\makebox(0,0)[b]{\smash{{{\SetFigFont{10}{14.4}{\rmdefault}{\mddefault}{\updefault}$\x$}}}}}
\put(3608,864){\makebox(0,0)[b]{\smash{{{\SetFigFont{10}{14.4}{\rmdefault}{\mddefault}{\updefault}$\y$}}}}}
\put(3233,39){\makebox(0,0)[b]{\smash{{{\SetFigFont{10}{14.4}{\rmdefault}{\mddefault}{\updefault}$y\st x$}}}}}
\end{picture}
}
  \end{center}
  \caption{Composition performed by VR.}
  \label{fig-VC}
\end{figure}

The preliminary step for our work is the Danos and Regnier's
construction of a confluent, local and asynchronous reduction of
$\lambda$-calculus, derived from a semantic setting based on a unique
type of move (simple enough to be easily mechanized). Their graph
reduction technique, namely VR, can be explained also as an efficient
way to compute the execution formula. The one and only reduction rule
is the composition of two edges in the graph as described in
Figure~\ref{fig-VC}. Whenever two edges of the virtual net are
composable (i.e. the product of their weights is non-null), VR derives
from them a new edge. The original edges are then marked by the
\emph{rest} of the composition, here denoted by weight within brackets.

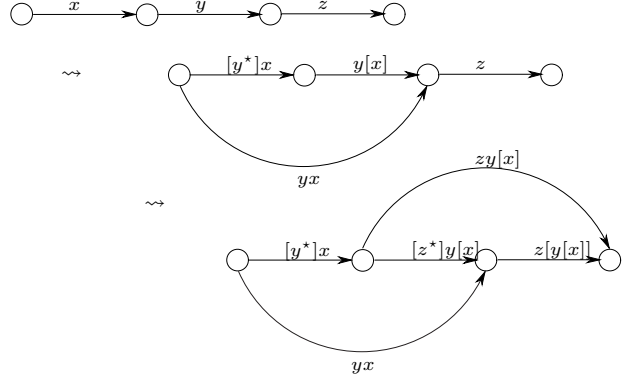
\begin{figure}[ht]
  \begin{center}
    \leavevmode
\setlength{\unitlength}{0.00055000in}
\begingroup\makeatletter\ifx\SetFigFont\undefined%
\gdef\SetFigFont#1#2#3#4#5{%
  \reset@font\fontsize{#1}{#2pt}%
  \fontfamily{#3}\fontseries{#4}\fontshape{#5}%
  \selectfont}%
\fi\endgroup%
{\renewcommand{\dashlinestretch}{30}
\begin{picture}(5834,3635)(0,-10)
\put(3354.000,1498.608){\arc{2569.527}{0.4164}{2.7252}}
\blacken\path(4501.773,858.341)(4529.000,979.000)(4448.196,885.348)(4491.189,903.991)(4501.773,858.341)
\put(4539.000,685.392){\arc{2569.527}{3.5580}{5.8668}}
\blacken\path(5633.196,1298.652)(5714.000,1205.000)(5686.773,1325.659)(5676.189,1280.009)(5633.196,1298.652)
\put(2170,1090){\ellipse{212}{212}}
\put(4540,1090){\ellipse{212}{212}}
\put(3365,1085){\ellipse{212}{212}}
\put(5720,1090){\ellipse{212}{212}}
\path(4652,1095)(5625,1095)
\blacken\path(5505.000,1065.000)(5625.000,1095.000)(5505.000,1125.000)(5541.000,1095.000)(5505.000,1065.000)
\path(2282,1090)(3255,1090)
\blacken\path(3135.000,1060.000)(3255.000,1090.000)(3135.000,1120.000)(3171.000,1090.000)(3135.000,1060.000)
\path(3477,1090)(4450,1090)
\blacken\path(4330.000,1060.000)(4450.000,1090.000)(4330.000,1120.000)(4366.000,1090.000)(4330.000,1060.000)
\put(4990,1140){\makebox(0,0)[lb]{\smash{{{\SetFigFont{7}{8.4}{\rmdefault}{\mddefault}{\updefault}$\z$}}}}}
\put(3255,45){\makebox(0,0)[lb]{\smash{{{\SetFigFont{7}{8.4}{\rmdefault}{\mddefault}{\updefault}$yx$}}}}}
\put(2615,1130){\makebox(0,0)[lb]{\smash{{{\SetFigFont{7}{8.4}{\rmdefault}{\mddefault}{\updefault}$\xs$}}}}}
\put(3825,1140){\makebox(0,0)[lb]{\smash{{{\SetFigFont{7}{8.4}{\rmdefault}{\mddefault}{\updefault}$\yss$}}}}}
\put(4430,2005){\makebox(0,0)[lb]{\smash{{{\SetFigFont{7}{8.4}{\rmdefault}{\mddefault}{\updefault}$z\ys$}}}}}
\put(2799.000,3265.608){\arc{2569.527}{0.4164}{2.7252}}
\blacken\path(3946.773,2625.341)(3974.000,2746.000)(3893.196,2652.348)(3936.189,2670.991)(3946.773,2625.341)
\put(114,3435){\ellipse{212}{212}}
\put(1309,3430){\ellipse{212}{212}}
\put(3664,3435){\ellipse{212}{212}}
\put(2484,3435){\ellipse{212}{212}}
\put(1615,2857){\ellipse{212}{212}}
\put(3985,2857){\ellipse{212}{212}}
\put(2810,2852){\ellipse{212}{212}}
\put(5165,2857){\ellipse{212}{212}}
\path(226,3435)(1199,3435)
\blacken\path(1079.000,3405.000)(1199.000,3435.000)(1079.000,3465.000)(1115.000,3435.000)(1079.000,3405.000)
\path(1421,3435)(2394,3435)
\blacken\path(2274.000,3405.000)(2394.000,3435.000)(2274.000,3465.000)(2310.000,3435.000)(2274.000,3405.000)
\path(2596,3440)(3569,3440)
\blacken\path(3449.000,3410.000)(3569.000,3440.000)(3449.000,3470.000)(3485.000,3440.000)(3449.000,3410.000)
\path(4097,2862)(5070,2862)
\blacken\path(4950.000,2832.000)(5070.000,2862.000)(4950.000,2892.000)(4986.000,2862.000)(4950.000,2832.000)
\path(1727,2857)(2700,2857)
\blacken\path(2580.000,2827.000)(2700.000,2857.000)(2580.000,2887.000)(2616.000,2857.000)(2580.000,2827.000)
\path(2922,2857)(3895,2857)
\blacken\path(3775.000,2827.000)(3895.000,2857.000)(3775.000,2887.000)(3811.000,2857.000)(3775.000,2827.000)
\put(559,3475){\makebox(0,0)[lb]{\smash{{{\SetFigFont{7}{8.4}{\rmdefault}{\mddefault}{\updefault}$x$}}}}}
\put(1769,3485){\makebox(0,0)[lb]{\smash{{{\SetFigFont{7}{8.4}{\rmdefault}{\mddefault}{\updefault}$y$}}}}}
\put(2934,3485){\makebox(0,0)[lb]{\smash{{{\SetFigFont{7}{8.4}{\rmdefault}{\mddefault}{\updefault}$z$}}}}}
\put(491,2825){\makebox(0,0)[lb]{\smash{{{\SetFigFont{7}{8.4}{\rmdefault}{\mddefault}{\updefault}$\leadsto$}}}}}
\put(1286,1589){\makebox(0,0)[lb]{\smash{{{\SetFigFont{7}{8.4}{\rmdefault}{\mddefault}{\updefault}$\leadsto$}}}}}
\put(4435,2907){\makebox(0,0)[lb]{\smash{{{\SetFigFont{7}{8.4}{\rmdefault}{\mddefault}{\updefault}$z$}}}}}
\put(2060,2897){\makebox(0,0)[lb]{\smash{{{\SetFigFont{7}{8.4}{\rmdefault}{\mddefault}{\updefault}$\xs$}}}}}
\put(3270,2907){\makebox(0,0)[lb]{\smash{{{\SetFigFont{7}{8.4}{\rmdefault}{\mddefault}{\updefault}$\ys$}}}}}
\put(2740,1816){\makebox(0,0)[lb]{\smash{{{\SetFigFont{7}{8.4}{\rmdefault}{\mddefault}{\updefault}$yx$}}}}}
\end{picture}
}
  \end{center}
  \caption{VR originating bars of bars.}
  \label{fig.barsofbars}
\end{figure}

The algebraic mechanism corresponding to the rest is called the
\emph{bar}; it was introduced in \cite{DanosRegnier93} to ensure 
the preservation of Girard's execution formula. Note that VR induces
bars of bars by definition; this is shown in
Figure~\ref{fig.barsofbars}. DVR, presented in \cite{DPR97}, was
designed in order to avoid bars of bars, thus allowing any
implementation to use simple data structures for representing edges.

\subsection{Geometry of Interaction}
\label{goi}

The basic geometrical construction consists of a directed graph with
weights in the dynamic algebra.  The most important point is that the
computation of Girard's ``Execution Formula'' is performed in a way
that appears to be the natural candidate for parallel computation. In
order to get a computational device from this graphical calculus, a
suitable strategy has been introduced: by means of the
\emph{combustion} strategy it was proved that not only the mechanism
of DVR computes the execution formula but also that it can do it in
the same way as Lamping's algorithm for sharing graphs.
  
The Geometry of Interaction basic step is the introduction of a
suitable algebraic structure, in view of the modeling of the dynamics
of the reduction. This structure can be thought as of the set of
partial one-to-one maps $u$ with composition.  The structure is then
enriched with partial inverses $u\st$, the codomain operation $\fin
u$, and the complementary of the codomain $\fil u$. Axioms for such a
structure are formally introduced below.

\begin{definition}\label{im}
An \emph{inverse monoid} (see \cite{Petrich84}), or for short an
\emph{im}, is a monoid with an unary function, the \emph{star},
denoted by $(.)\st$, with
\begin{eqnarray}
  (uv)\st&=&v\st u\st,\\
  (u\st)\st&=&u,\\
  uu\st u&=&u,\\
  uu\st vv\st &=& vv\st uu\st.
\end{eqnarray}
\end{definition} 
We denote by $\fin u$ the idempotent $uu\st$. With this notation the
last equation becomes $\fin u\fin v = \fin v\fin u$ and the one before
becomes $\fin u u=u$. 

\begin{definition}
A \emph{bar inverse monoid}, or for short a \emph{bim}, is an im with
a zero, denoted by 0, and an unary function, the \emph{bar}, denoted
by $[.]$, with
\begin{eqnarray}
  \fil 1 = 0&\mbox{and}&\fil 0 = 1,\\
  u\fil v   &=& \fil {uv} u.
\end{eqnarray}
\end{definition}

Bim's axioms entail:
\begin{enumerate}
\item\label{puno}  $\fil{u}u=0$, in fact $u\fil{1}=\fil{u \, 1}u$ thus $0 = \fil{u}u$;
\item  $\fil u\fil u=\fil u$, in fact $\fil{u}\fil{u} = \fil{\fil{u}u}\fil{u} =
\fil{0}\fil{u} = \fil{u}$;
\item $\fil{u}\st = \fil{u}$, 
$\fil{u}\st\fil{u} = \fil{u}\st \fil{u}\fil{u}\st \fil{u} =
                    \fil{u}\st \fil{u} \fil{u}\st\fil{u} \st\fil{u} =
                    \fil{u}\st\fil{u}\fil{u}\fil{u}\st \fil{u}\st \fil{u} =\\
                    \fil{u}\st\fil{u}\fil{u}\st\fil{u} \fil{u} \fil{u}\st =
                    \fil{u}\st\fil{u} \fil{u} \fil{u}\st =
                    \fil{u}\st\fil{u} \fil{u}\st = \fil{u}\st
$
then we have  
$\fil{u} = \fil{u} \fil{u}\st \fil{u} 
          = \fil{u} \fil{u}\st \fil{u} \fil{u}\st \fil{u}  
          = \fil{u} \fil{u}\st  \fil{u}\st \fil{u} 
          = \fil{u}\st \fil{u} \fil{u} \fil{u}\st
          = \fil{u}\st  \fil{u} \fil{u}\st
          = \fil{u}\st$;

\item and  $vu=0$ iff $v\fil{u}=v$. In fact 
$v\fil{u} = \fil{vu}v = \fil{0} v = v$, on the other hand
if  $v\fil{u}=v$ then  $v\fil{u}u=vu$ but $\fil{u}u=0$ thus
$v\fil{u}u=0 =vu$.
\end{enumerate}

Now we give the construction of the free bim generated by a given im.
So let $S$ be an im, and $\Z[S]$ denote the free contracted algebra
over $S$ with coefficients in $\Z$ (the ring of integers). In other
words $\Z[S]$ is the algebra of maps from $S$ to $\Z$ with finitely
many non-zero values. In other words $\Z[S]$ is the algebra of linear
combinations over $S$ with coefficients in $\Z$.

For any such linear combination, $s=\sum n_is_i$, define
\begin{equation}\label{sept}
s\st=\sum n_is_i\st,\qquad
\fil{s}=1-\fin s=1-ss\st.
\end{equation}

Define the \emph{complementary closure} of $S$ in $\Z[S]$, denoted by
$\fil S$, as the monoid generated in $\Z[S]$ by the union of $S$ and
$\{1-\fin u,\ u\in S\}$.

\begin{proposition}
$\fil S$ is an inverse monoid with $(.)\st$ defined as in (\ref{sept}).
\end{proposition}

\begin{proof}
The proof is a straightforward calculation. In fact we have that for
every element $ s \in \fil S$ can be written as a combination
\begin{equation}\label{combfil}
s = s_1 \fil{u_1} s_2 \dots \fil{u_{n}} s_{n+1}.
\end{equation}

Let us introduce the length $|s|$ of an element $s \in \fil{S}$ as the
smallest $n$ such that (\ref{combfil}). 

Now we prove that properties (1)-(4) in Definition \ref{im} hold 
for $\fil S$. 

\begin{enumerate}

\item First we prove that for every
$u,v \in \fil{S}$ we have $(u v )\st = v\st u\st$, by double induction
on the lengths of $u$ and $v$.

If $|u|=0$ and $|v|=0$ then $u,v \in S$ and
the property holds by definition, because $S$ is an inverse monoid.

Let be $|u| = 0$ and for every $v$ such that $|v|\le n$ the property holds,
then we prove that for every $v'$ such that $|v'| = n+1$,
$(u v')\st = v'\st u\st$.

Let be $v'= v [u_{n+1}] s_{n+2}$ and $u = s$, then 
\begin{align*}
(uv')\st &= (s v [u_{n+1}] s_{n+2})\st=\\
         &= (s v (1 - \fin{u_{n+1}}) s_{n+2})\st =\\
         &= (s v s_{n+2} - s v \fin{u_{n+1}} s_{n+2})\st =\\
         &= (s v s_{n+2})\st - (s v \fin{u_{n+1}} s_{n+2})\st = \\
         &=s_{n+2}\st v\st s\st  -  s_{n+2}\st\fin{u_{n+1}} v\st s\st =\\
         &=s_{n+2}\st [u_{n+1}] v\st s\st = v'\st u\st,
\end{align*}
in fact $|v's_{n+2}|= n$ and from $\fin{u_{n+1}} s_{n+2} \in S$
we have $|v' \fin{u_{n+1}} s_{n+2}|=n$; thus by induction hypothesis
 $(s v s_{n+2})\st = s_{n+2}\st v\st s\st $ and
$(s v' \fin{u_{n+1}} s_{n+2})\st$ $=s_{n+2}\st\fin{u_{n+1}} v'\st s\st$.
Now suppose the property holds for any $u$ such that $|u|\le n$ and
for every $v \in \fil{S}$. We show that it holds for $u'\in \fil{S}$
such that $|u'| = n+1$.
\begin{align*}
(u'v)\st 
         &= (u [u_{n+1}] s_{n+2} v)\st=\\
         &= (u s_{n+2}v - u u_{n+1}u_{n+1}\st s_{n+2} v )\st=\\
         &= ( u s_{n+2}v)\st - ( u u_{n+1}u_{n+1}\st s_{n+2} v )\st,
\end{align*}
we have $|u s_{n+2}|=|u|=n$ and
$|u u_{n+1}u_{n+1}\st s_{n+2}|=n$
by induction hypothesis, we have $( u s_{n+2}v)\st=
 v\st s_{n+2}\st u\st$ and 
$$( u u_{n+1}u_{n+1}\st s_{n+2} v )\st=v\st s_{n+2}\st u_{n+1}u_{n+1}\st u\st,$$ 
thus
\begin{align*}
(u'v)\st
 &= v\st s_{n+2}\st u\st - v\st s_{n+2}\st u_{n+1}u_{n+1}\st u\st=\\
         &= v\st s_{n+2}\st [u_{n+1}] u\st =\\
         & = v\st (u [u_{n+1}] s_{n+2})\st = v\st u'\st.
\end{align*}

\item 
Now we prove that $u\st\st = u$. Again by induction on the length $|u|$.
It is clear that if $|u|=0$ then $u\in S$ and $u\st\st = u $ by definition
of inverse monoid.

Suppose that the for every $u \in  \fil{S}$ such that $|u|\le n$ we have
$u\st\st = u$, and consider $u'$ such that $|u'| = n+1$. We may write
$u' = u \fil{u_{n+1}} s_{n+2}$, then
we have $u'\st = (u \fil{u_{n+1}} s_{n+2})\st$ by the previous proof
we have $ (u \fil{u_{n+1}} s_{n+2})\st = s_{n+2}\st \fil{u_{n+1}} u\st$
and 
\begin{align*}
u'\st\st =& ( s_{n+2}\st \fil{u_{n+1}} u\st)\st=\\
         =& u\st\st \fil{u_{n+1}} s_{n+2}\st\st=\\
         =& u \fil{u_{n+1}} s_{n+2},
\end{align*}
since by induction hypothesis $u\st\st = u$.


\item 
Let us prove that for every $u \in \fil{S}$ we have $uu\st u =u$.
Case $|u|=0$, implies $u\in S$ and  follows from the definition of $S$.

Suppose, $uu\st u = u$ for every $u \in \fil{S}$ such that
$|u| \le n$. Consider $u' = u \fil{u_{n+1}} s_{n+2} $ such that
$|u|=n$ and $|u'|=n+1$. Then
\begin{align*}
u'u'\st u' &= u \fil{u_{n+1}} s_{n+2} (u \fil{u_{n+1}} s_{n+2})\st u \fil{u_{n+1}} s_{n+2} =\\
&=u \fil{u_{n+1}} s_{n+2} s_{n+2}\st \fil{u_{n+1}} u\st u \fil{u_{n+1}} s_{n+2} =\\
&=u (1- \fin{u_{n+1}}) s_{n+2} s_{n+2}\st (1- \fin{u_{n+1}})  u\st u (1- \fin{u_{n+1}}) s_{n+2} =\\
&=u  s_{n+2} s_{n+2}\st u\st u  s_{n+2} \\
& \qquad -u \fin{u_{n+1}} s_{n+2} s_{n+2}\st u\st u  s_{n+2} \\
& \qquad -u  s_{n+2} s_{n+2}\st \fin{u_{n+1}} u\st u  s_{n+2} 
         -u  s_{n+2} s_{n+2}\st u\st u \fin{u_{n+1}} s_{n+2}  \\
& \qquad +u  \fin{u_{n+1}} s_{n+2} s_{n+2}\st \fin{u_{n+1}} u\st u  s_{n+2} \\
& \qquad +u  \fin{u_{n+1}} s_{n+2} s_{n+2}\st u\st u \fin{u_{n+1}} s_{n+2} \\
& \qquad +u  s_{n+2} s_{n+2}\st\fin{u_{n+1}} u\st u \fin{u_{n+1}} s_{n+2} \\
& \qquad -u  \fin{u_{n+1}} s_{n+2} s_{n+2}\st \fin{u_{n+1}} u\st u \fin{u_{n+1}} s_{n+2} = \\
\end{align*}
\begin{align*}
&=u  s_{n+2} -
u  s_{n+2} s_{n+2}\st u\st u s_{n+2} s_{n+2}\st \fin{u_{n+1}} s_{n+2} \\
& \qquad -u  s_{n+2} s_{n+2}\st \fin{u_{n+1}} u\st u  s_{n+2} \\
& \qquad -u  s_{n+2} s_{n+2}\st u\st u \fin{u_{n+1}} s_{n+2}  \\
& \qquad +u  \fin{u_{n+1}} s_{n+2} s_{n+2}\st \fin{u_{n+1}} u\st u  s_{n+2} \\
& \qquad +u  \fin{u_{n+1}} s_{n+2} s_{n+2}\st u\st u \fin{u_{n+1}} s_{n+2} \\
& \qquad  +u  s_{n+2} s_{n+2}\st\fin{u_{n+1}} u\st u \fin{u_{n+1}} s_{n+2} \\
& \qquad  -u  \fin{u_{n+1}} s_{n+2} s_{n+2}\st \fin{u_{n+1}} u\st u \fin{u_{n+1}} s_{n+2} =\\
&=u  s_{n+2} -
u  s_{n+2} s_{n+2}\st \fin{u_{n+1}} s_{n+2} \\
& \qquad -u  s_{n+2} s_{n+2}\st  u\st u  s_{n+2} s_{n+2}\st \fin{u_{n+1}} s_{n+2}\\ 
& \qquad -u  s_{n+2} s_{n+2}\st u\st u \fin{u_{n+1}} s_{n+2}  \\
& \qquad +u s_{n+2} s_{n+2}\st u\st u s_{n+2} s_{n+2}\st \fin{u_{n+1}}  s_{n+2} \\
& \qquad +u s_{n+2} s_{n+2}\st u\st u s_{n+2} s_{n+2}\st \fin{u_{n+1}} \fin{u_{n+1}} s_{n+2} \\
& \qquad  +u  s_{n+2} s_{n+2}\st\fin{u_{n+1}} u\st u \fin{u_{n+1}} s_{n+2}\\ 
& \qquad  -u  \fin{u_{n+1}} s_{n+2} s_{n+2}\st \fin{u_{n+1}} u\st u \fin{u_{n+1}} s_{n+2} =\\
&=u  s_{n+2} - 3 u  \fin{u_{n+1}} s_{n+2} + 3 u  \fin{u_{n+1}}  s_{n+2} 
   -u \fin{u_{n+1}} s_{n+2} =\\
&=us_{n+2}-u\fin{u_{n+1}}s_{n+2}=\\ 
&= u\fil{u_{n+1}}s_{n+2}= \\
&= u',
\end{align*}
we used 
$ u s_{n+2} s_{n+2}\st u\st u s_{n+2} = u s_{n+2}$
 by applying the induction hypothesis to $u$.


\item 
We prove that for any $u,v \in \fil{S}$,
$uu\st vv\st = vv\st uu\st$. Let us fix $|u|=0$ and
let us show the property for any $v$ by induction on 
$|v|$.

So $|u|=0$ and if $|v|=0$ 
then $u,v \in S$ and there is nothing to prove.

Let us apply induction hypothesis, and for a 
fixed $n$ suppose the property holds for any $|u|=0$ and $v$ such
that $|v|\le n$. We then prove the property holds for any 
$v'$ such that $|v'|= n+1$, in this case we may 
suppose that $v' = v\fil{u_{n+1}} s_{n+2}$,
moreover by definition of $\fil{u_{n+1}} = 1 - u_{n+1} u_{n+1}\st$ and by distribution,  
\begin{align*}
 ss\st v\fil{u_{n+1}} s_{n+2} s_{n+2}\st \fil{u_{n+1}}v\st  
&= ss\st v\fil{u_{n+1}}\fil{u_{n+1}} s_{n+2} s_{n+2}\st v\st= \\
&= ss\st v\fil{u_{n+1}} s_{n+2} s_{n+2}\st v\st=\\
&= ss\st v(1-\fin{u_{n+1}}) s_{n+2} s_{n+2}\st v\st=\\
&= ss\st v s_{n+2} s_{n+2}\st v\st
-ss\st v \fin{u_{n+1}} s_{n+2} s_{n+2}\st v\st.
\end{align*}

In the last expression we have $|v s_{n+2}| = n$, thus 
from the basis of the induction
$$ss\st v s_{n+2} s_{n+2} v\st= v s_{n+2} s_{n+2} v\st ss\st.$$  
In a similar way $\fin{u_{n+1}}, s_{n+2} \in S$ 
and so
\begin{align*}
v\fin{u_{n+1}}s_{n+2} s_{n+2}\st v\st 
&= v\fin{u_{n+1}}\fin{u_{n+1}}s_{n+2} s_{n+2}\st v\st=\\
&= v\fin{u_{n+1}}s_{n+2} s_{n+2}\st \fin{u_{n+1}}v\st,
\end{align*}
and 
$|v\fin{u_{n+1}}s_{n+2}|=n$, thus
$$ss\st v\fin{u_{n+1}}s_{n+2} s_{n+2}\st v\st =
v\fin{u_{n+1}}s_{n+2} s_{n+2}\st \fin{u_{n+1}} v\st ss\st$$
by induction hypothesis.

Now we complete the proof by an induction on $|u|$, 
suppose that  the property holds for any $u$ such that 
$|u|\le n$ and for any $v\in \fil{S}$ and  
consider $u' = u\fil{u_{n+1}} s_{n+2}$,
then
\begin{align*}
u\fil{u_{n+1}} s_{n+2} s_{n+2}\st \fil{u_{n+1}}&u\st v v\st=\\
&= u\fil{u_{n+1}} \fil{u_{n+1}} s_{n+2} s_{n+2}\st u\st v v\st=\\
&= u\fil{u_{n+1}} s_{n+2} s_{n+2}\st u\st v v\st=\\
&= u s_{n+2} s_{n+2}\st u\st v v\st-
u\fin{u_{n+1}} s_{n+2} s_{n+2}\st u\st v v\st=\\
&= u s_{n+2} s_{n+2}\st u\st v v\st-
u\fin{u_{n+1}} s_{n+2} s_{n+2}\st \fin{u_{n+1}} u\st v v\st,
\end{align*}
since $\fil{u_{n+1}} s_{n+2} \in S$
we have $| u\fil{u_{n+1}} s_{n+2}| = n$
and  by induction hypothesis we obtain
\begin{align*}
v v\st u s_{n+2} s_{n+2}\st u\st &-v v\st u\fin{u_{n+1}} s_{n+2} s_{n+2}\st \fin{u_{n+1}} u\st =\\
&=v v\st u\fil{u_{n+1}} s_{n+2} s_{n+2}\st \fil{u_{n+1}}u\st.
\end{align*}
\end{enumerate}
\end{proof}


\begin{definition}
Define the \emph{bar closure} of $S$, denoted by $\fil S_\om$,
to be the im obtained by $\om$ iterations of the complementary
closure, that is: let $S_0=S$ and $S_{n+1}=\fil{S_n}$ then $\fil
S_\om= \bigcup_{n\geq0}S_n$.  
\end{definition}

\begin{proposition}
$\fil S_\om$ is a bar inverse monoid with $\fil{.}$ defined as above.
\end{proposition}

\begin{proof}
For every $u,v \in \fil S_\om$ there exists $n$ s.t. $u,v \in \fil
S_n$ so $u\,v$, $u\st$ and $[u]$ belong to $\fil S_n$ and so to $\fil
S_\om$; for the same reason bim's axioms are satisfied.
\end{proof}

\begin{definition}
The monoid $\Ls$ of the Geometry of Interaction is the free monoid with a
morphism $!(.)$, an involution $(.)\st$ and a zero, generated by $p$,
$q$, and a family $W=(w_{i})_i$ of exponential generators such that for
any $u\in\Ls$:
\begin{eqnarray}
  x\st y &=& \delta_{xy} \qquad\mbox{for $x,y=p,q,w_{i}$,} \label{annhy}\\
   !(u)w_{i} &=& w_{i}!^{e_{i}} (u), \label{commy}
\end{eqnarray}
\noindent
where $e_{i}$ is an integer associated with $w_{i}$ called the 
\emph{lift} of $w_{i}$, $i$ is called the \emph{name} of $w_i$ and 
we will often write $w_{i,e(i)}$ to explicitly note the lift of the
generator.

Equations (\ref{annhy}) will be called of \emph{annihilation} and
(\ref{commy}) are called equations of \emph{swapping}.
\end{definition}

Orienting the equations (\ref{annhy}-\ref{commy}) from left to right,
one gets a rewriting system which is terminating and confluent. The
non-zero normal forms, known as \emph{stable forms}, are the terms
$ab\st$ where $a$ and $b$ are \emph{positive} (i.e. written without
$\st$s). The fact that all non-zero terms are equal to such an $ab\st$
form is referred to as the ``$ab\st$ property''.  From this, one
easily gets that the word problem is decidable and that \Ls\ is an
inverse monoid.

Every computation, from now on, will take place in the bar closure of
$\Ls$ in \KLs, which we denote by $\fil\Ls_\om$. Since, as said, this
is a bim, results in \cite{DanosRegnier93}, which were stated and
proved for any bar inverse monoid, apply with no further ado. Note
that equalities in $\fil\Ls_\om$ and in \KLs\ are also decidable by
rewriting to stable form.

Set $\fil{b_1,\dots,b_n}=1-b_1b_1\st-\cdots-b_nb_n\st$; 
$\fil{b_1,\dots,b_n}$ is an idempotent iff the
$b_i$'s are \emph{orthogonal} that is, $\fin{b_i}\fin{b_j}=0$.  

\begin{lemma}[(superposition)] 
Let $a$, $b$ and $c$ be positive monomials in \Ls\ 
such that $\fin{a}\fin{b}$, $\fin{b}\fin{c}$ and 
$\fin{a}\fin{c}\neq 0$, then $\fin{a}\fin{b}\fin{c}\neq 0$.
\end{lemma}

\begin{proof}
 See \cite{DPR97}.
\end{proof}

\begin{definition}
Let a \emph{weight} on a directed graph be a functor $\W$ from the
directed graph's involutive category of paths to $\fil\Ls_\om$.
\end{definition}

Most of the time, we will simply write $\phi$ for $\W(\phi)$ to ease
the reading of definitions and proofs.

We will say that $\alpha$ \emph{coincides} with $\beta$ or equivalently 
that $\alpha$ and $\beta$ are coincident if they have the same target node.

An edge $\beta$ is called a \emph{counter-edge} of $\alpha$ along
$\tau$ if $\beta \not= \alpha$ and $\tau$ is a directed path from the
$\alpha$'s target to the $\beta$'s one, not ending with $\beta$,
such that $\fin{\al}\fin{\tau\st\beta}\not=0$.

Two coincident counter-edges $\alpha$ and $\beta$ are said to be
\emph{composable} (i.e. $\fin{\alpha}\fin{\beta}\not = 0$
or equivalently they are reciprocally counter edges along the empty
path).

\begin{definition}
A \emph{straight path} is a path that contains no sub-path of the 
form $\phi\phi\st$, i.e. that never bounces back in the same edge.
\end{definition}

A weighted directed graph is said to be \emph{split} if any three
coincident paths of length one $\phi_1$, $\phi_2$ and $\phi_3$ are
such that $\fin{\phi_1}\fin{\phi_2}\fin{\phi_3}=0$; it is said to be
\emph{square-free} if for any straight path $\phi$, $\phi\phi=0$.  

\begin{definition}
A weighted directed graph is said to be a \emph{virtual net} if it is split and
square-free.
\end{definition}

Splitness can be rephrased as: any three paths $\phi_1$, $\phi_2$
and $\phi_3$ such that none is prefix of another are such that
$$\fin{\phi_1}\fin{\phi_2}\fin{\phi_3}=0.$$
                
\subsection{Directed Virtual Reduction}
\label{pippo}

\begin{definition}
A {\it directed virtual net\/} $R$ is an acyclic virtual net 
such that for each edge $\al$:

{\bf A.} $\al=\fil{b_1,\dots,b_n}a$, where $a$, $b_1$, \dots, $b_n$
are positive monomials of \Ls. We will denote by $\al\pl$ the weight
of $\al$ without its \emph{filter} $\fil{b_1,\dots, b_n}$ that is, the
monomial $a$.

{\bf B.} for any $i\neq j$ and for any two counter-edges $\ba_1$,
$\ba_2$ of $\al$ along $\tau_1$, $\tau_2$
\[\begin{array}{l@{{}=0\qquad}l}
\fin{b_i}\fin{b_j}&\cl 0R\al{b_i}{b_j}\\
\fin{b_i}\fin{\tau_1\st\ba_1\pl}&\cl 1R\al{b_i}{\tau_1\st\ba_1}\\
\fin{\tau_{1}\st\ba_1\pl} \fin{\tau_{2}\st\ba_2\pl}&
\cl 2R\al{\tau_{1}\st\ba_1}{\tau_{2}\st\ba_2}
\end{array}\]
\end{definition}

\begin{figure}[ht]
\def\a{$\scriptstyle\fil{b_1,\dots,b_n}a$}
\def\b{$\scriptstyle\fil{a_1,\dots,a_m}b$}
\def\c{$\scriptstyle\fil{b_1,\dots,b_n,b}a$}
\def\d{$\scriptstyle\fil{a_1,\dots,a_m,a}b$}
\def\e{$\scriptstyle b'$}
\def\f{$\scriptstyle a'$}
\def\y{$\leadsto$}
\centerline{
\setlength{\unitlength}{0.00055833in}
\begingroup\makeatletter\ifx\SetFigFont\undefined%
\gdef\SetFigFont#1#2#3#4#5{%
  \reset@font\fontsize{#1}{#2pt}%
  \fontfamily{#3}\fontseries{#4}\fontshape{#5}%
  \selectfont}%
\fi\endgroup%
{\renewcommand{\dashlinestretch}{30}
\begin{picture}(5291,2115)(0,-10)
\put(3195.136,1881.579){\arc{3525.401}{1.5351}{2.7211}}
\path(1665.953,1067.621)(1586.000,1162.000)(1612.133,1041.099)
\put(3548.864,1881.579){\arc{3525.401}{0.4205}{1.6065}}
\path(5131.867,1041.099)(5158.000,1162.000)(5078.047,1067.621)
\put(3377,1261){\ellipse{212}{212}}
\put(1577,1261){\ellipse{212}{212}}
\put(5177,1261){\ellipse{212}{212}}
\put(3369,113){\ellipse{212}{212}}
\put(1914,1860){\ellipse{212}{212}}
\put(114,1860){\ellipse{212}{212}}
\put(3714,1860){\ellipse{212}{212}}
\path(5064,1261)(3486,1260)
\path(3605.981,1290.076)(3486.000,1260.000)(3606.019,1230.076)
\path(1689,1261)(3261,1260)
\path(3140.981,1230.076)(3261.000,1260.000)(3141.019,1290.076)
\path(3601,1860)(2025,1860)
\path(2145.000,1890.000)(2025.000,1860.000)(2145.000,1830.000)
\path(226,1860)(1806,1863)
\path(1686.057,1832.772)(1806.000,1863.000)(1685.943,1892.772)
\put(1727,1336){\makebox(0,0)[lb]{\smash{{{\SetFigFont{7}{8.4}{\rmdefault}{\mddefault}{\updefault}$[b_1,\dots,b_n,b]a$}}}}}
\put(3602,1336){\makebox(0,0)[lb]{\smash{{{\SetFigFont{7}{8.4}{\rmdefault}{\mddefault}{\updefault}$[a_1,\dots,a_n,a]b$}}}}}
\put(264,1935){\makebox(0,0)[lb]{\smash{{{\SetFigFont{7}{8.4}{\rmdefault}{\mddefault}{\updefault}$[b_1,\dots,b_n]a$}}}}}
\put(2139,1935){\makebox(0,0)[lb]{\smash{{{\SetFigFont{7}{8.4}{\rmdefault}{\mddefault}{\updefault}$[a_1,\dots,a_n]b$}}}}}
\put(776,1280){\makebox(0,0)[lb]{\smash{{{\SetFigFont{7}{8.4}{\rmdefault}{\mddefault}{\updefault}$\leadsto$}}}}}
\put(1866,363){\makebox(0,0)[lb]{\smash{{{\SetFigFont{7}{8.4}{\rmdefault}{\mddefault}{\updefault}$b'$}}}}}
\put(4641,363){\makebox(0,0)[lb]{\smash{{{\SetFigFont{7}{8.4}{\rmdefault}{\mddefault}{\updefault}$a'$}}}}}
\end{picture}
}
}
\caption{Composition performed by DVR.}\label{fig.dirvired}
\end{figure}
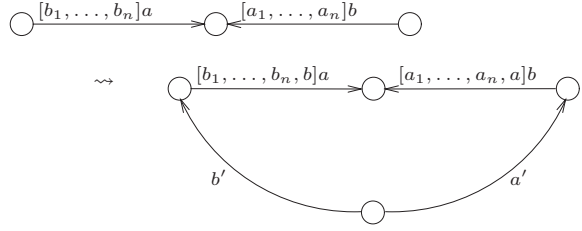

Given two coincident counter-edges $\alpha$ and $\beta$,
with weights $\fil{b_1,\dots,b_n}a$ and $\fil{a_1,\dots,a_m}b$, then
DVR originates a new node and two new edges linking that node to the
sources of $\alpha$ and $\beta$. These new edges have, respectively,
weights $b'$ and $a'$ where $a'{b'}\st$ is the stable form of $b\st
a$; this is shown in Figure \ref{fig.dirvired}.

Note that new edges produced by a step of reduction have positive
weights so that the resulting computation of the execution formula is
more appealing for the implementation, as opposed to VR, by the fact
that bars are not propagated on residuals.

\begin{definition}
Given two composable edges $\alpha$ and $\beta$, the two edges
$\alpha'$ and $\beta'$ generated by one step of DVR are called
\emph{residuals} of $\alpha$ and $\beta$ respectively.
We will denote these residuals by $$\dvrstep(\alpha,\beta)=(\beta',\alpha').$$
\end{definition}

\begin{lemma}[(augmentation)]  \label{augmentation}
Let $R$ be a directed virtual net and $\ga_2$ be a counter-edge of
$\ga_1$ along $\tau$ in $R$, then $$
\fin{\ga_1}\fin{\tau\st\ga_2}= \fin{\ga_1\pl}\fin{\tau\pl\st\ga_2\pl}\neq0,
$$
or equivalently
$$
\ga_1\st\tau\st\ga_2=\ga_1\pl\st\tau\pl\st\ga_2\pl\neq0.
$$
\end{lemma}

\begin{proof} See \cite{DPR97}.
\end{proof}

In \cite{DPR97}, it has been proved that DVR is sound w.r.t. Girard's
execution formula:
\begin{proposition}[(Invariance)] The execution for\-mu\-la 
is an invariant 
of DVR.  \end{proposition}
\begin{proof} See \cite{DPR97}.
\end{proof}

\subsection{Optimization Rules}\label{optrules}

In this section we prove two properties in order to make DVR more
effective, which are also exploited while developing  the implementation. These properties
are immediate consequences of the orthogonality conditions satisfied
by virtual nets and  help to gain effectiveness in the computation and to
increase the intrinsic parallelism.

\begin{definition} \label{ghost}
Given a directed virtual net $R$, a \emph{total} edge $\alpha$ is any edge with
at most one counter-edge $\beta$ that is,
$\beta$ is the only edge such that
$$\fin{\alpha}\fin{\tau\st_{\alpha\beta}\beta}\not=0.$$
In this case  we say that $\alpha$ is \emph{total w.r.t. $\beta$}; if
 $\alpha$ has no counter edge, it is called \emph{ghost}.
\end{definition}

This relation is not symmetric: for any two coincident edges $\alpha$
and $\beta$ such that $\alpha$ is total w.r.t. $\beta$, we observe
that it is possible that $\beta$ is not total w.r.t. $\alpha$, in
fact: suppose $\beta = 1$, then any $\gamma$ coincident with $\beta$
is composable with $\beta$ (thus $\beta$ cannot be total w.r.t.
$\alpha$) but $\alpha$ is not composable with $\gamma$ otherwise it
would contradict the splitness condition, thus $\alpha$ is total
w.r.t. $\beta$.

\begin{proposition}\label{matching}
Given two composable edges $\alpha$ and $\beta$ such that
$\dvrstep(\alpha,\beta)=(1,\alpha')$, then $\alpha$ is total
w.r.t. $\beta$.
\end{proposition}

\begin{proof} In order to get a contradiction suppose that there exists a
counter-edge $\gamma$ of $\alpha$ along the directed path $\tau$.
Suppose $\tau$ is the empty path, therefore $\gamma$ coincides with
$\alpha$ and $\alpha\st \gamma \not=0$ so that
$\dvrstep(\alpha,\gamma)=(\gamma',\alpha'')$; in this case we compute
$\fin{\gamma}\fin{\alpha}\fin{\beta}$ and we have $\gamma
\gamma\st\alpha \alpha\st \beta \beta\st
=\gamma\alpha''\gamma'\st\alpha'\st\beta\st \not=0$ since it is a
stable form, and we get a contradiction with the splitness condition.

If $\tau$ is not the empty path, we can apply the same argument to the
residual $\gamma'$ of the reduction sequence along the directed path
$\tau$, and derive the property by lemma \ref{augmentation} applied to
$\alpha$ and $\gamma$.
\end{proof}

\begin{proposition}\label{m2}
Given three coincident edges $\alpha, \beta, \gamma$ such that
$$\dvrstep(\alpha , \beta)=(\beta',1)$$ with $\alpha$ and $\gamma$
composable, then the residual of $\gamma$ is not composable with $\beta'$.
\end{proposition}

\begin{proof} Suppose $\gamma'\st \beta' \not= 0$ so this product
has a stable form, let say $\tilde{\beta'}\tilde{\gamma'}\st$, then
$\fin{\gamma}\fin{\alpha}\fin{\beta}=\gamma\gamma\st\alpha\alpha\st\beta\beta\st=\gamma\alpha''\gamma'\st\beta'\beta\st=\gamma\alpha''\tilde{\beta'}\tilde{\gamma'}\st\beta\st\not=0$
and we get a contradiction with the splitness condition.
\end{proof}

\begin{corollary} (soundness of the optimization of one)
If $\dvrstep(\alpha, \beta)=(1,\alpha')$, then no further composable
edge $\gamma$ can give $1$ as residual of the composition with
$\alpha$.
\end{corollary}

\begin{proof} The two residuals in the source of $\alpha$  have weight $1$ 
so that they are composable and this is a contradiction with the
proposition \ref{m2}.
\end{proof}

\begin{figure}[ht]
\centerline{
\begin{picture}(0,0)%
\epsfig{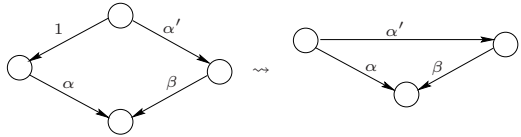}%
\end{picture}%
\setlength{\unitlength}{2605sp}%
\begingroup\makeatletter\ifx\SetFigFont\undefined%
\gdef\SetFigFont#1#2#3#4#5{%
  \reset@font\fontsize{#1}{#2pt}%
  \fontfamily{#3}\fontseries{#4}\fontshape{#5}%
  \selectfont}%
\fi\endgroup%
\begin{picture}(4874,1261)(0,-413)
\put(476,563){\makebox(0,0)[lb]{\smash{\SetFigFont{6}{7.2}{\rmdefault}{\mddefault}{\updefault}$1$}}}
\put(1510,572){\makebox(0,0)[lb]{\smash{\SetFigFont{6}{7.2}{\rmdefault}{\mddefault}{\updefault}$\alpha'$}}}
\put(2361,176){\makebox(0,0)[lb]{\smash{\SetFigFont{6}{7.2}{\rmdefault}{\mddefault}{\updefault}$\leadsto$}}}
\put(3630,542){\makebox(0,0)[lb]{\smash{\SetFigFont{6}{7.2}{\rmdefault}{\mddefault}{\updefault}$\alpha'$}}}
\put(555, 49){\makebox(0,0)[lb]{\smash{\SetFigFont{6}{7.2}{\rmdefault}{\mddefault}{\updefault}$\alpha$}}}
\put(1540, 54){\makebox(0,0)[lb]{\smash{\SetFigFont{6}{7.2}{\rmdefault}{\mddefault}{\updefault}$\beta$}}}
\put(3435,202){\makebox(0,0)[lb]{\smash{\SetFigFont{6}{7.2}{\rmdefault}{\mddefault}{\updefault}$\alpha$}}}
\put(4070,201){\makebox(0,0)[lb]{\smash{\SetFigFont{6}{7.2}{\rmdefault}{\mddefault}{\updefault}$\beta$}}}
\end{picture}
}
  \caption{An optimization rule.}
  \label{oneopt}
\end{figure}

This corollary allows an optimization rule, in fact the configuration
produced by the DVR step $\dvrstep(\alpha,\beta)=(1,\alpha')$ acts as
a compound operator: the edge with weight 1 is there just to say that
all the coincident edges have to be transferred on the source of the
edge $\alpha'$, so we propose to transform this configuration by
removing the edge $\beta'$ with weight 1 and using the edge $\alpha'$
for linking the target of $\beta'$ and the target of $\alpha'$ (see
Figure \ref{oneopt}).

Now we will prove another property of DVR, which 
states that when two edges $\alpha'_1$
and $\alpha'_2$ are residuals of directed virtual reduction of  $\alpha_1$ and $\alpha_2$ against the same edge $\beta$, they
are coincident on the source of $\beta$ (evident by definition of
a DVR step), but are not composable because of splitness:
\begin{proposition}\label{orthogonalsystems} 
Given an edge $\beta$ composable with two coincident edges $\alpha_1$
and $\alpha_2$ we have that $\fin{\alpha_1'} \fin{\alpha_2'} =0 $, where
$\alpha_1'$ is the residual of $\alpha_1$ and  
$\alpha_2'$ is the residual of $\alpha_2$.
\end{proposition}

\begin{proof}
Suppose $\fin{\alpha_1'} \fin{\alpha_2'} \not=0 $
then we have, 
$\alpha_1' \alpha_1'\st \alpha_2' \alpha_2'\st$
and so
$\alpha_1'\st \alpha'_2 = \alpha_2'' \alpha''_1\st \not=0$.

By augmentation lemma we have
$$\fin{\alpha_1} \fin{\beta} \fin{\alpha_2} =
\fin{\alpha_1^+} \fin{\beta^+} \fin{\alpha_2^+},$$
this is 
$$\alpha_1 \alpha_1\st \beta\beta\st \alpha_2\alpha_2\st$$
and by reduction we obtain
$$\alpha_1 \beta' \alpha_2'' \alpha''_1\st \beta''\st \alpha_2\st$$
and so
this is a non null stable form so that it is different from zero.
\end{proof}

Property \ref{orthogonalsystems} allows the implementation of
another optimization rule. Specifically, we know that every new node
$v$ created after a DVR step is the source of only two edges, say
$\beta_1$ and $\beta_2$, therefore all the edges coincident in $v$ can
be separated in two sets: the residuals of a DVR step involving
$\beta_1$ and the residuals of a DVR step involving $\beta_2$.  Each
edge in a set is orthogonal to any edge in the same set, therefore
there is no need to perform DVR steps between edges belonging to the
same set since the composition will actually produce a null result.

\subsection{Translation of Sharing Graphs into Directed Virtual Nets}
\label{trans}

In order to solve the problem of the pairing of duplication operators
Gonthier et al. added to the sharing graphs a local level structure.
Each operator is decorated with an integer tag that specifies the
\emph{level} at which it lives. Furthermore in order to manage these levels a
set of control operators is required.

More precisely sharing graphs are non-oriented graphs built from the
indexed nodes represented in Figure \ref{sharing}. These nodes are
called sharing operators and distinguished in two groups. The first
group includes the operators in the original Lamping's work:
application, and abstraction; the second one is constituted by a
family of nodes of the same kind (the so called \emph{muxes})
accordingly to the following definition:
\begin{figure}[ht]
  \begin{center}
    \leavevmode
\begin{picture}(0,0)%
\epsfig{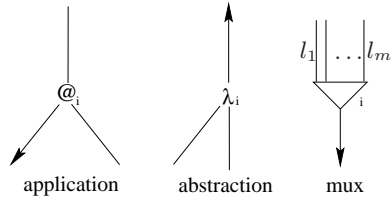}%
\end{picture}%
\setlength{\unitlength}{2960sp}%
\begingroup\makeatletter\ifx\SetFigFont\undefined%
\gdef\SetFigFont#1#2#3#4#5{%
  \reset@font\fontsize{#1}{#2pt}%
  \fontfamily{#3}\fontseries{#4}\fontshape{#5}%
  \selectfont}%
\fi\endgroup%
\begin{picture}(3006,1634)(492,-1428)
\put(2922,-239){\makebox(0,0)[lb]{\smash{\SetFigFont{9}{10.8}{\rmdefault}{\mddefault}{\updefault}$l_1$}}}
\put(3213,-242){\makebox(0,0)[lb]{\smash{\SetFigFont{9}{10.8}{\rmdefault}{\mddefault}{\updefault}$\dots$}}}
\put(3483,-236){\makebox(0,0)[lb]{\smash{\SetFigFont{9}{10.8}{\rmdefault}{\mddefault}{\updefault}$l_m$}}}
\end{picture}
\end{center}
  \caption{Sharing graph operators.}
  \label{sharing}
\end{figure}
\begin{definition}
A node \emph{mux} or multiplexer is a node with an arbitrary number
of premises each one having a \emph{name} $n$ and a 
\emph{lift} $l_n$; like the other nodes, muxes have an index of level $i$.
\end{definition}

The translation of a sharing graph with muxes is defined by induction:
\begin{definition}\label{trnodes} A sharing graph $M$ with root $x$ and context $y_1,
\dots , y_n$
is translated into a directed virtual net in the following way

\centerline{
\begin{picture}(0,0)%
\epsfig{file=generic.term.pstex}%
\end{picture}%
\setlength{\unitlength}{2960sp}%
\begingroup\makeatletter\ifx\SetFigFont\undefined%
\gdef\SetFigFont#1#2#3#4#5{%
  \reset@font\fontsize{#1}{#2pt}%
  \fontfamily{#3}\fontseries{#4}\fontshape{#5}%
  \selectfont}%
\fi\endgroup%
\begin{picture}(898,955)(107,-265)
\put(107,-104){\makebox(0,0)[lb]{\smash{\SetFigFont{8}{9.6}{\rmdefault}{\mddefault}{\updefault}$y_1$}}}
\put(478,-216){\makebox(0,0)[lb]{\smash{\SetFigFont{8}{9.6}{\rmdefault}{\mddefault}{\updefault}$y_2$}}}
\put(948,-107){\makebox(0,0)[lb]{\smash{\SetFigFont{8}{9.6}{\rmdefault}{\mddefault}{\updefault}$y_n$}}}
\put(680,555){\makebox(0,0)[lb]{\smash{\SetFigFont{8}{9.6}{\rmdefault}{\mddefault}{\updefault}$x$}}}
\put(547,179){\makebox(0,0)[lb]{\smash{\SetFigFont{8}{9.6}{\rmdefault}{\mddefault}{\updefault}$M$}}}
\put(544,-103){\makebox(0,0)[lb]{\smash{\SetFigFont{8}{9.6}{\rmdefault}{\mddefault}{\updefault}$\dots$}}}
\end{picture}
}

where bullets indicates
ports of a sharing graph,
\begin{itemize}
\item If $x$ is a link between two ports with no node

\centerline{
\begin{picture}(0,0)%
\epsfig{file=empty.pstex}%
\end{picture}%
\setlength{\unitlength}{2960sp}%
\begingroup\makeatletter\ifx\SetFigFont\undefined%
\gdef\SetFigFont#1#2#3#4#5{%
  \reset@font\fontsize{#1}{#2pt}%
  \fontfamily{#3}\fontseries{#4}\fontshape{#5}%
  \selectfont}%
\fi\endgroup%
\begin{picture}(1908,496)(743,256)
\put(2651,398){\makebox(0,0)[lb]{\smash{\SetFigFont{6}{7.2}{\rmdefault}{\mddefault}{\updefault}$1$}}}
\put(808,471){\makebox(0,0)[lb]{\smash{\SetFigFont{6}{7.2}{\rmdefault}{\mddefault}{\updefault}$x$}}}
\put(1345,476){\makebox(0,0)[lb]{\smash{\SetFigFont{6}{7.2}{\rmdefault}{\mddefault}{\updefault}$\leadsto$}}}
\end{picture}
}

\item If $x$ is a port of an abstraction

\centerline{
\begin{picture}(0,0)%
\epsfig{file=abstraction.pstex}%
\end{picture}%
\setlength{\unitlength}{2960sp}%
\begingroup\makeatletter\ifx\SetFigFont\undefined%
\gdef\SetFigFont#1#2#3#4#5{%
  \reset@font\fontsize{#1}{#2pt}%
  \fontfamily{#3}\fontseries{#4}\fontshape{#5}%
  \selectfont}%
\fi\endgroup%
\begin{picture}(3149,1262)(488,-456)
\put(1028,-254){\makebox(0,0)[lb]{\smash{\SetFigFont{6}{7.2}{\rmdefault}{\mddefault}{\updefault}$\dots$}}}
\put(1031, 28){\makebox(0,0)[lb]{\smash{\SetFigFont{6}{7.2}{\rmdefault}{\mddefault}{\updefault}$M$}}}
\put(1149,625){\makebox(0,0)[lb]{\smash{\SetFigFont{6}{7.2}{\rmdefault}{\mddefault}{\updefault}$x$}}}
\put(962,-367){\makebox(0,0)[lb]{\smash{\SetFigFont{6}{7.2}{\rmdefault}{\mddefault}{\updefault}$y_2$}}}
\put(1432,-258){\makebox(0,0)[lb]{\smash{\SetFigFont{6}{7.2}{\rmdefault}{\mddefault}{\updefault}$y_n$}}}
\put(1073,504){\makebox(0,0)[lb]{\smash{\SetFigFont{6}{7.2}{\rmdefault}{\mddefault}{\updefault}$\lambda$}}}
\put(668,-252){\makebox(0,0)[lb]{\smash{\SetFigFont{6}{7.2}{\rmdefault}{\mddefault}{\updefault}$y_1$}}}
\put(1911, 79){\makebox(0,0)[lb]{\smash{\SetFigFont{6}{7.2}{\rmdefault}{\mddefault}{\updefault}$\leadsto$}}}
\put(3176,-228){\makebox(0,0)[lb]{\smash{\SetFigFont{6}{7.2}{\rmdefault}{\mddefault}{\updefault}$\dots$}}}
\put(3179, 54){\makebox(0,0)[lb]{\smash{\SetFigFont{6}{7.2}{\rmdefault}{\mddefault}{\updefault}$M$}}}
\put(3110,-341){\makebox(0,0)[lb]{\smash{\SetFigFont{6}{7.2}{\rmdefault}{\mddefault}{\updefault}$y_2$}}}
\put(3580,-232){\makebox(0,0)[lb]{\smash{\SetFigFont{6}{7.2}{\rmdefault}{\mddefault}{\updefault}$y_n$}}}
\put(2816,-226){\makebox(0,0)[lb]{\smash{\SetFigFont{6}{7.2}{\rmdefault}{\mddefault}{\updefault}$y_1$}}}
\put(1171,419){\makebox(0,0)[lb]{\smash{\SetFigFont{6}{7.2}{\rmdefault}{\mddefault}{\updefault}$i$}}}
\put(3358,485){\makebox(0,0)[lb]{\smash{\SetFigFont{6}{7.2}{\rmdefault}{\mddefault}{\updefault}$!^iq$}}}
\put(2773,551){\makebox(0,0)[lb]{\smash{\SetFigFont{6}{7.2}{\rmdefault}{\mddefault}{\updefault}$!^ip$}}}
\end{picture}
}
\item If $x$ is a port of an application

\centerline{
\begin{picture}(0,0)%
\includegraphics{application.pstex}%
\end{picture}%
\setlength{\unitlength}{2960sp}%
\begingroup\makeatletter\ifx\SetFigFont\undefined%
\gdef\SetFigFont#1#2#3#4#5{%
  \reset@font\fontsize{#1}{#2pt}%
  \fontfamily{#3}\fontseries{#4}\fontshape{#5}%
  \selectfont}%
\fi\endgroup%
\begin{picture}(4071,1489)(181,-451)
\put(1171,419){\makebox(0,0)[lb]{\smash{\SetFigFont{6}{7.2}{\rmdefault}{\mddefault}{\updefault}
$i$
}}}
\put(1149,625){\makebox(0,0)[lb]{\smash{\SetFigFont{6}{7.2}{\rmdefault}{\mddefault}{\updefault}
$x$
}}}
\put(916,-252){\makebox(0,0)[lb]{\smash{\SetFigFont{6}{7.2}{\rmdefault}{\mddefault}{\updefault}
$y_n$
}}}
\put(1462,-302){\makebox(0,0)[lb]{\smash{\SetFigFont{6}{7.2}{\rmdefault}{\mddefault}{\updefault}
$\dots$
}}}
\put(1465,-20){\makebox(0,0)[lb]{\smash{\SetFigFont{6}{7.2}{\rmdefault}{\mddefault}{\updefault}
$M$
}}}
\put(1396,-415){\makebox(0,0)[lb]{\smash{\SetFigFont{6}{7.2}{\rmdefault}{\mddefault}{\updefault}
$y'_2$
}}}
\put(1102,-300){\makebox(0,0)[lb]{\smash{\SetFigFont{6}{7.2}{\rmdefault}{\mddefault}{\updefault}
$y'_1$
}}}
\put(541,-296){\makebox(0,0)[lb]{\smash{\SetFigFont{6}{7.2}{\rmdefault}{\mddefault}{\updefault}
$\dots$
}}}
\put(544,-14){\makebox(0,0)[lb]{\smash{\SetFigFont{6}{7.2}{\rmdefault}{\mddefault}{\updefault}
$N$
}}}
\put(475,-409){\makebox(0,0)[lb]{\smash{\SetFigFont{6}{7.2}{\rmdefault}{\mddefault}{\updefault}
$y_2$
}}}
\put(181,-294){\makebox(0,0)[lb]{\smash{\SetFigFont{6}{7.2}{\rmdefault}{\mddefault}{\updefault}
$y_1$
}}}
\put(1058,504){\makebox(0,0)[lb]{\smash{\SetFigFont{6}{7.2}{\rmdefault}{\mddefault}{\updefault}
@
}}}
\put(4181,-277){\makebox(0,0)[lb]{\smash{\SetFigFont{6}{7.2}{\rmdefault}{\mddefault}{\updefault}
$y'_m$
}}}
\put(3245,-232){\makebox(0,0)[lb]{\smash{\SetFigFont{6}{7.2}{\rmdefault}{\mddefault}{\updefault}
$y_n$
}}}
\put(3791,-282){\makebox(0,0)[lb]{\smash{\SetFigFont{6}{7.2}{\rmdefault}{\mddefault}{\updefault}
$\dots$
}}}
\put(3794,  0){\makebox(0,0)[lb]{\smash{\SetFigFont{6}{7.2}{\rmdefault}{\mddefault}{\updefault}
$M$
}}}
\put(3725,-395){\makebox(0,0)[lb]{\smash{\SetFigFont{6}{7.2}{\rmdefault}{\mddefault}{\updefault}
$y'_2$
}}}
\put(3431,-280){\makebox(0,0)[lb]{\smash{\SetFigFont{6}{7.2}{\rmdefault}{\mddefault}{\updefault}
$y'_1$
}}}
\put(2870,-276){\makebox(0,0)[lb]{\smash{\SetFigFont{6}{7.2}{\rmdefault}{\mddefault}{\updefault}
$\dots$
}}}
\put(2873,  6){\makebox(0,0)[lb]{\smash{\SetFigFont{6}{7.2}{\rmdefault}{\mddefault}{\updefault}
$N$
}}}
\put(2804,-389){\makebox(0,0)[lb]{\smash{\SetFigFont{6}{7.2}{\rmdefault}{\mddefault}{\updefault}
$y_2$
}}}
\put(2510,-274){\makebox(0,0)[lb]{\smash{\SetFigFont{6}{7.2}{\rmdefault}{\mddefault}{\updefault}
$y_1$
}}}
\put(1852,-297){\makebox(0,0)[lb]{\smash{\SetFigFont{6}{7.2}{\rmdefault}{\mddefault}{\updefault}
$y'_m$
}}}
\put(2024, 15){\makebox(0,0)[lb]{\smash{\SetFigFont{6}{7.2}{\rmdefault}{\mddefault}{\updefault}
$\leadsto$
}}}
\put(2946,625){\makebox(0,0)[lb]{\smash{\SetFigFont{6}{7.2}{\rmdefault}{\mddefault}{\updefault}
$!^iq$
}}}
\put(3400,488){\makebox(0,0)[lb]{\smash{\SetFigFont{6}{7.2}{\rmdefault}{\mddefault}{\updefault}
$!^ip$
}}}
\put(3475,833){\makebox(0,0)[lb]{\smash{\SetFigFont{6}{7.2}{\rmdefault}{\mddefault}{\updefault}
1
}}}
\end{picture}
}
\item If $y_i$ is a port of a mux

\centerline{
\begin{picture}(0,0)%
\epsfig{file=muxt.pstex}%
\end{picture}%
\setlength{\unitlength}{2960sp}%
\begingroup\makeatletter\ifx\SetFigFont\undefined%
\gdef\SetFigFont#1#2#3#4#5{%
  \reset@font\fontsize{#1}{#2pt}%
  \fontfamily{#3}\fontseries{#4}\fontshape{#5}%
  \selectfont}%
\fi\endgroup%
\begin{picture}(2919,1406)(625,-766)
\put(1031, 28){\makebox(0,0)[lb]{\smash{\SetFigFont{6}{7.2}{\rmdefault}{\mddefault}{\updefault}$M$}}}
\put(1432,-258){\makebox(0,0)[lb]{\smash{\SetFigFont{6}{7.2}{\rmdefault}{\mddefault}{\updefault}$y_n$}}}
\put(1161,366){\makebox(0,0)[lb]{\smash{\SetFigFont{6}{7.2}{\rmdefault}{\mddefault}{\updefault}$x$}}}
\put(1130,-628){\makebox(0,0)[lb]{\smash{\SetFigFont{6}{7.2}{\rmdefault}{\mddefault}{\updefault}$y_i$}}}
\put(709,-350){\makebox(0,0)[lb]{\smash{\SetFigFont{6}{7.2}{\rmdefault}{\mddefault}{\updefault}$\dots$}}}
\put(906,-266){\makebox(0,0)[lb]{\smash{\SetFigFont{6}{7.2}{\rmdefault}{\mddefault}{\updefault}$y'_1$}}}
\put(1147,-265){\makebox(0,0)[lb]{\smash{\SetFigFont{6}{7.2}{\rmdefault}{\mddefault}{\updefault}$y'_m$}}}
\put(1024,-315){\makebox(0,0)[lb]{\smash{\SetFigFont{6}{7.2}{\rmdefault}{\mddefault}{\updefault}$\dots$}}}
\put(788,-264){\makebox(0,0)[lb]{\smash{\SetFigFont{6}{7.2}{\rmdefault}{\mddefault}{\updefault}$\dots$}}}
\put(625,-242){\makebox(0,0)[lb]{\smash{\SetFigFont{6}{7.2}{\rmdefault}{\mddefault}{\updefault}$y_1$}}}
\put(1276,-254){\makebox(0,0)[lb]{\smash{\SetFigFont{6}{7.2}{\rmdefault}{\mddefault}{\updefault}$\dots$}}}
\put(1911, 79){\makebox(0,0)[lb]{\smash{\SetFigFont{6}{7.2}{\rmdefault}{\mddefault}{\updefault}$\leadsto$}}}
\put(3086,137){\makebox(0,0)[lb]{\smash{\SetFigFont{6}{7.2}{\rmdefault}{\mddefault}{\updefault}$M$}}}
\put(3487,-149){\makebox(0,0)[lb]{\smash{\SetFigFont{6}{7.2}{\rmdefault}{\mddefault}{\updefault}$y_n$}}}
\put(3216,475){\makebox(0,0)[lb]{\smash{\SetFigFont{6}{7.2}{\rmdefault}{\mddefault}{\updefault}$x$}}}
\put(2764,-241){\makebox(0,0)[lb]{\smash{\SetFigFont{6}{7.2}{\rmdefault}{\mddefault}{\updefault}$\dots$}}}
\put(2961,-157){\makebox(0,0)[lb]{\smash{\SetFigFont{6}{7.2}{\rmdefault}{\mddefault}{\updefault}$y'_1$}}}
\put(3202,-156){\makebox(0,0)[lb]{\smash{\SetFigFont{6}{7.2}{\rmdefault}{\mddefault}{\updefault}$y'_m$}}}
\put(3079,-206){\makebox(0,0)[lb]{\smash{\SetFigFont{6}{7.2}{\rmdefault}{\mddefault}{\updefault}$\dots$}}}
\put(2843,-155){\makebox(0,0)[lb]{\smash{\SetFigFont{6}{7.2}{\rmdefault}{\mddefault}{\updefault}$\dots$}}}
\put(2680,-133){\makebox(0,0)[lb]{\smash{\SetFigFont{6}{7.2}{\rmdefault}{\mddefault}{\updefault}$y_1$}}}
\put(3331,-145){\makebox(0,0)[lb]{\smash{\SetFigFont{6}{7.2}{\rmdefault}{\mddefault}{\updefault}$\dots$}}}
\put(3022,-431){\makebox(0,0)[lb]{\smash{\SetFigFont{5}{6.0}{\rmdefault}{\mddefault}{\updefault}$\dots$}}}
\put(2726,-402){\makebox(0,0)[lb]{\smash{\SetFigFont{5}{6.0}{\rmdefault}{\mddefault}{\updefault}$!^i w_{1 l_1}$}}}
\put(3265,-407){\makebox(0,0)[lb]{\smash{\SetFigFont{5}{6.0}{\rmdefault}{\mddefault}{\updefault}$!^i w_{m l_m}$}}}
\end{picture}
}
\end{itemize}
\end{definition}

\begin{figure}[ht]
\centerline{
\begin{picture}(0,0)%
\epsfig{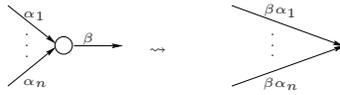}%
\end{picture}%
\setlength{\unitlength}{2960sp}%
\begingroup\makeatletter\ifx\SetFigFont\undefined%
\gdef\SetFigFont#1#2#3#4#5{%
  \reset@font\fontsize{#1}{#2pt}%
  \fontfamily{#3}\fontseries{#4}\fontshape{#5}%
  \selectfont}%
\fi\endgroup%
\begin{picture}(2829,734)(64,-606)
\put(1276,-286){\makebox(0,0)[lb]{\smash{\SetFigFont{5}{6.0}{\rmdefault}{\mddefault}{\updefault}$\leadsto$}}}
\put(226, 14){\makebox(0,0)[lb]{\smash{\SetFigFont{5}{6.0}{\rmdefault}{\mddefault}{\updefault}$\alpha_1$}}}
\put(234,-319){\makebox(0,0)[lb]{\smash{\SetFigFont{5}{6.0}{\rmdefault}{\mddefault}{\updefault}$\vdots$}}}
\put(222,-553){\makebox(0,0)[lb]{\smash{\SetFigFont{5}{6.0}{\rmdefault}{\mddefault}{\updefault}$\alpha_n$}}}
\put(2239, 44){\makebox(0,0)[lb]{\smash{\SetFigFont{5}{6.0}{\rmdefault}{\mddefault}{\updefault}$\beta\alpha_1$}}}
\put(2233,-575){\makebox(0,0)[lb]{\smash{\SetFigFont{5}{6.0}{\rmdefault}{\mddefault}{\updefault}$\beta\alpha_n$}}}
\put(2293,-316){\makebox(0,0)[lb]{\smash{\SetFigFont{5}{6.0}{\rmdefault}{\mddefault}{\updefault}$\vdots$}}}
\put(721,-196){\makebox(0,0)[lb]{\smash{\SetFigFont{5}{6.0}{\rmdefault}{\mddefault}{\updefault}$\beta$}}}
\end{picture}
}
\caption{Amalgamation of edges.}
\label{famal} 
\end{figure}

When the nodes introduced by the translation present the configuration
described in the next definition they are reduced by amalgamating
edges as in Figure \ref{famal}.
\begin{definition}\label{tramal} A node with $n$
coincident edges $\alpha_1 , \dots,$ $\alpha_n$ and an edge $\beta$
with source the target of the $\alpha_i$'s is erased and all the
$\alpha_i$'s are replaced by edges $\alpha'_i$ where the source of
$\alpha'_i$ is the source of $\alpha_i$, the target of $\alpha'_i$ is
the target of $\beta$ and the weight of $\alpha'_i$ is $\alpha_i
\beta$.  
\end{definition}

\begin{figure}[ht]
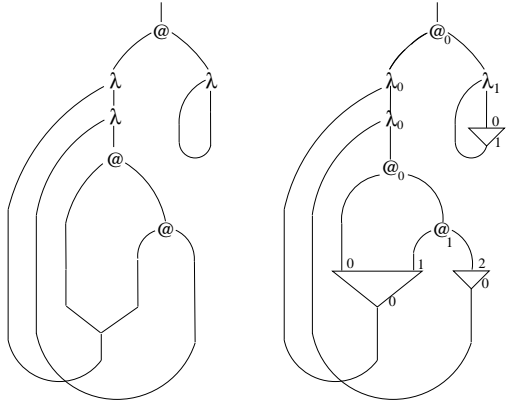

  \begin{center}
    \leavevmode
\begin{picture}(0,0)%
\epsfig{file=2Il.pstex}%
\end{picture}%
\setlength{\unitlength}{2960sp}%
\begingroup\makeatletter\ifx\SetFigFont\undefined%
\gdef\SetFigFont#1#2#3#4#5{%
  \reset@font\fontsize{#1}{#2pt}%
  \fontfamily{#3}\fontseries{#4}\fontshape{#5}%
  \selectfont}%
\fi\endgroup%
\begin{picture}(1720,3348)(1879,-3247)
\end{picture}
\hspace{2em}
\begin{picture}(0,0)%
\epsfig{file=2I.bis.pstex}%
\end{picture}%
\setlength{\unitlength}{2960sp}%
\begingroup\makeatletter\ifx\SetFigFont\undefined%
\gdef\SetFigFont#1#2#3#4#5{%
  \reset@font\fontsize{#1}{#2pt}%
  \fontfamily{#3}\fontseries{#4}\fontshape{#5}%
  \selectfont}%
\fi\endgroup%
\begin{picture}(1869,3348)(1879,-3247)
\end{picture}
  \end{center}
  \caption{Representation of a $\lambda$-term.}
  \label{2I}
\end{figure}

\begin{figure}[ht]
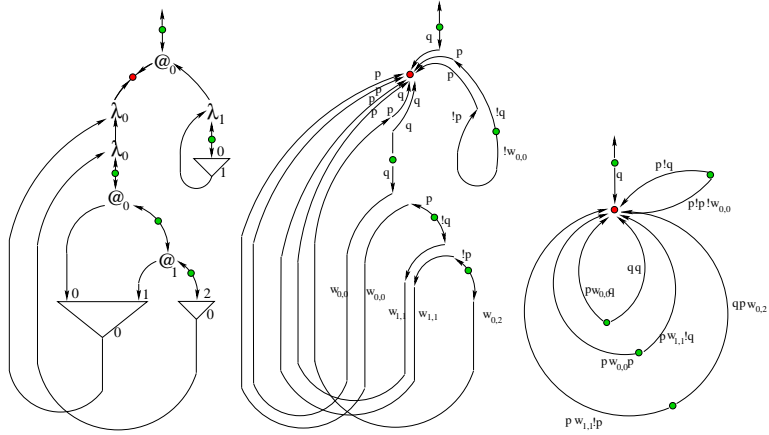

  \begin{center}
    \leavevmode
\begin{picture}(0,0)%
\epsfig{file=2Istep1.bis.pstex}%
\end{picture}%
\setlength{\unitlength}{2960sp}%
\begingroup\makeatletter\ifx\SetFigFont\undefined%
\gdef\SetFigFont#1#2#3#4#5{%
  \reset@font\fontsize{#1}{#2pt}%
  \fontfamily{#3}\fontseries{#4}\fontshape{#5}%
  \selectfont}%
\fi\endgroup%
\begin{picture}(1869,3549)(1879,-3247)
\end{picture}
\begin{picture}(0,0)%
\epsfig{file=2Istep2bis.pstex}%
\end{picture}%
\setlength{\unitlength}{2960sp}%
\begingroup\makeatletter\ifx\SetFigFont\undefined%
\gdef\SetFigFont#1#2#3#4#5{%
  \reset@font\fontsize{#1}{#2pt}%
  \fontfamily{#3}\fontseries{#4}\fontshape{#5}%
  \selectfont}%
\fi\endgroup%
\begin{picture}(2307,3603)(1503,-3254)
\end{picture}
\begin{picture}(0,0)%
\epsfig{file=2Idvr.pstex}%
\end{picture}%
\setlength{\unitlength}{2960sp}%
\begingroup\makeatletter\ifx\SetFigFont\undefined%
\gdef\SetFigFont#1#2#3#4#5{%
  \reset@font\fontsize{#1}{#2pt}%
  \fontfamily{#3}\fontseries{#4}\fontshape{#5}%
  \selectfont}%
\fi\endgroup%
\begin{picture}(1948,2485)(2161,-2132)
\end{picture}
  \end{center}
  \caption{Encoding of a sharing graph into a directed virtual net.}
  \label{dvrex}
\end{figure}

With the help of an example we show how to change a $\lambda$-term into a
directed virtual net.  In Figure \ref{2I}, starting from the syntactic
graph of the $\lambda$-term representing the Church numeral 2 applied
to the identity that is, by using Krivine's notation, $(\lambda f
\lambda x (f)(f)x)\lambda x x$, we obtain a sharing graph by adding
the control operators, expressed in the multiplexer syntax, and
annotating each node by level indexes.
Then edges are oriented, unfolded and labeled with monomials in $\Ls$
in accord to the rules expressed by Definition \ref{trnodes} and Definition
\ref{tramal}.  Last step consists of grouping together arrows going in
the same direction, the result of this operation is a directed virtual
net, see Figure \ref{dvrex}.

Let us check that the obtained net is indeed a directed virtual net:
\begin{itemize}
\item it is obviously a directed graph with no circuits, 
\item square-freeness can be proved by induction on the translation 
of $\lambda$-terms,
\item splitness has to be verified for all the triples of coincident edges;
as an example we explicitly test the splitness condition for three
coincident edges:
$$\fin{q}\fin{qpw_{02}}\fin{qq}=
qq\st qpw_{02}w_{02}p\st q\st qqq\st q\st =
qpw_{02}w_{02}p\st qq\st q\st=0
$$
because of $p\st q=0$; the rest of this verification
is left to the reader. 
\end{itemize}

\section{Half Combustion Strategy}
\label{hc}

In \cite{DPR97}, a strategy called \emph{combustion} is presented in
order to organize DVR in such a way that no filter must be kept.  This
strategy works on {\em full} directed virtual nets that are directed
virtual nets where each edge is either ghost (see
Definition~\ref{ghost}) or has a positive weight.

Since a ghost edge is an edge for which no more compositions will
occur, sources of ghost edges never receive residual edges of ghost
edges, thus let us define the (out-)\emph{valence} of a node as the
number of non-ghost edges having that node as source.

The combustion strategy of a full net starts from a node $v$ of
valence zero (i.e. with no future incoming edge or equivalently having
only ghost outgoing edges) and composes all the pairs of coincident
counter-edges on $v$ as an atomic action. Using the combustion
strategy we can give up filters because after the composition is
performed, all those edges become ghost edges.

From the point of view of a parallel implementation, the drawback of
this strategy is that the composition of the coincident counter-edges
can be started only when a node becomes of valence zero.  More
specifically, in case many processes are used to perform DVR (recall
this is desirable anytime we want to fully exploit the computing power
of parallel or distributed systems equipped with a large number of
processors), we might incur the risk that, at a given time instant,
only a subset of those processes host nodes of valence zero. In such a
case, all the other processes cannot simultaneously proceed with DVR
steps (i.e. they need to wait until some node they host becomes of
valence zero), thus limiting the degree of parallelism while
performing the reduction.

We define below the HC strategy that like combustion does not require
to keep filters and, in addition, allows the composition to be
performed even on nodes having valence greater than zero, thus
allowing high degree of parallelism. HC relies on the following notion
of {\em semifull} directed virtual net which is a generalization of
the notion of full directed virtual net.

\begin{figure}[ht]
  \begin{center}
    \leavevmode
    \setlength{\unitlength}{0.00055000in}
\begingroup\makeatletter\ifx\SetFigFont\undefined%
\gdef\SetFigFont#1#2#3#4#5{%
  \reset@font\fontsize{#1}{#2pt}%
  \fontfamily{#3}\fontseries{#4}\fontshape{#5}%
  \selectfont}%
\fi\endgroup%
{\renewcommand{\dashlinestretch}{30}
\begin{picture}(4059,2122)(0,-10)
\put(2603,1502){\ellipse{300}{300}}
\put(2603,1950){\ellipse{300}{300}}
\put(158,1544){\ellipse{300}{300}}
\put(1396,892){\ellipse{302}{302}}
\put(158,1087){\ellipse{300}{300}}
\put(169,385){\ellipse{310}{310}}
\put(2610,149){\ellipse{284}{284}}
\path(285,1465)(1245,947)
\blacken\path(1125.147,977.582)(1245.000,947.000)(1153.639,1030.386)(1171.075,986.889)(1125.147,977.582)
\blacken\path(1653.730,787.335)(1538.000,831.000)(1619.566,738.011)(1607.054,783.171)(1653.730,787.335)
\path(1538,831)(2475,182)
\blacken\path(1635.290,1036.544)(1545.000,952.000)(1664.453,984.108)(1618.410,992.828)(1635.290,1036.544)
\path(1545,952)(2453,1457)
\blacken\path(1572.595,1109.257)(1503.000,1007.000)(1612.528,1064.476)(1565.693,1062.907)(1572.595,1109.257)
\path(1503,1007)(2473,1872)
\path(308,1089)(1238,902)
\blacken\path(1114.441,896.244)(1238.000,902.000)(1126.269,955.067)(1155.648,918.559)(1114.441,896.244)
\path(315,400)(1253,805)
\blacken\path(1154.723,729.890)(1253.000,805.000)(1130.939,784.975)(1175.881,771.703)(1154.723,729.890)
\put(2213,1206){\makebox(0,0)[lb]{\smash{{{\SetFigFont{7}{8.4}{\rmdefault}{\mddefault}{\updefault}$[a_{21},\dots,a_{2j_2}]b_2$}}}}}
\put(2243,636){\makebox(0,0)[lb]{\smash{{{\SetFigFont{7}{8.4}{\rmdefault}{\mddefault}{\updefault}$\vdots$}}}}}
\put(627,1297){\makebox(0,0)[lb]{\smash{{{\SetFigFont{7}{8.4}{\rmdefault}{\mddefault}{\updefault}$a_1$}}}}}
\put(642,667){\makebox(0,0)[lb]{\smash{{{\SetFigFont{7}{8.4}{\rmdefault}{\mddefault}{\updefault}$a_m$}}}}}
\put(627,1057){\makebox(0,0)[lb]{\smash{{{\SetFigFont{7}{8.4}{\rmdefault}{\mddefault}{\updefault}$a_2$}}}}}
\put(900,1768){\makebox(0,0)[lb]{\smash{{{\SetFigFont{7}{8.4}{\rmdefault}{\mddefault}{\updefault}$[a_{11},\dots,a_{1j_1}]b_1$}}}}}
\put(136,643){\makebox(0,0)[lb]{\smash{{{\SetFigFont{7}{8.4}{\rmdefault}{\mddefault}{\updefault}$\vdots$}}}}}
\put(818,240){\makebox(0,0)[lb]{\smash{{{\SetFigFont{7}{8.4}{\rmdefault}{\mddefault}{\updefault}$[a_{n1},\dots,a_{nj_n}]b_n$}}}}}
\end{picture}
}
  \end{center}
  \caption{A node in a semifull directed virtual net.}
  \label{semifull}
\end{figure}
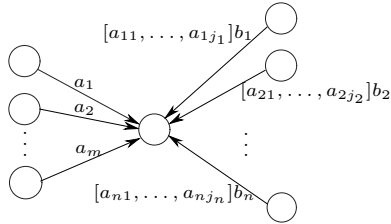

Let us call \emph{semifull} directed virtual net a directed virtual
net in which each edge either is weighted by a positive monomial
(i.e. its weight has no filter) or all its coincident counter-edges
are weighted by a positive monomial (i.e. it can be composed
exclusively with edges having a positive weight).  An example of a
node in a semifull directed virtual net is shown in Figure
\ref{semifull}. In this example, the coincident counter-edges of 
edges with weight $[a_{i1},\dots, a_{ij_1}]b_i$ are among those
edges weighted with $a_1$, $a_2$, ..., $a_m$.

Below we give the definition and provide the soundness of the HC
strategy. 

\begin{definition} \label{HCdef}
Given a composable edge $\alpha$ with positive weight in a semifull
directed virtual net $R$, we have to consider two cases:
\begin{enumerate}

\item if $\alpha$ has no non-positive coincident counter-edge and 
a positive one $\beta$, then the \emph{half combustion strategy} (HC)
performs the composition of $\beta$ with $\alpha$ and possibly with
every non-positive edge composable with $\beta$;
\item if the set $\{\beta_1, \dots , \beta_n\}$ of non-positive edges 
composable with $\alpha$ is non-empty then HC performs all the
possible compositions of $\alpha$ with the $\beta_i$s.
\end{enumerate}
\end{definition}

\begin{proposition} 
If $R'$ is obtained from the directed virtual net $R$ 
by the HC strategy and $R$ is semifull then so is $R'$.
\end{proposition}

\begin{figure}[ht]
  \begin{center}
    \leavevmode
    \setlength{\unitlength}{0.00055000in}
\begingroup\makeatletter\ifx\SetFigFont\undefined%
\gdef\SetFigFont#1#2#3#4#5{%
  \reset@font\fontsize{#1}{#2pt}%
  \fontfamily{#3}\fontseries{#4}\fontshape{#5}%
  \selectfont}%
\fi\endgroup%
{\renewcommand{\dashlinestretch}{30}
\begin{picture}(4296,2495)(0,-10)
\put(2643,1875){\ellipse{300}{300}}
\put(2643,2323){\ellipse{300}{300}}
\put(1436,1265){\ellipse{302}{302}}
\put(2650,522){\ellipse{284}{284}}
\put(158,1279){\ellipse{300}{300}}
\blacken\path(1693.730,1160.335)(1578.000,1204.000)(1659.566,1111.011)(1647.054,1156.171)(1693.730,1160.335)
\path(1578,1204)(2515,555)
\blacken\path(1675.290,1409.544)(1585.000,1325.000)(1704.453,1357.108)(1658.410,1365.828)(1675.290,1409.544)
\path(1585,1325)(2493,1830)
\blacken\path(1612.595,1482.257)(1543.000,1380.000)(1652.528,1437.476)(1605.693,1435.907)(1612.595,1482.257)
\path(1543,1380)(2513,2245)
\path(318,1280)(1268,1280)
\blacken\path(1148.000,1250.000)(1268.000,1280.000)(1148.000,1310.000)(1184.000,1280.000)(1148.000,1250.000)
\put(243,1425){\makebox(0,0)[lb]{\smash{{{\SetFigFont{7}{8.4}{\rmdefault}{\mddefault}{\updefault}$[b_1,\dots,b_n]a$}}}}}
\put(168,1650){\makebox(0,0)[lb]{\smash{{{\SetFigFont{7}{8.4}{\rmdefault}{\mddefault}{\updefault}$\vdots$}}}}}
\put(168,750){\makebox(0,0)[lb]{\smash{{{\SetFigFont{7}{8.4}{\rmdefault}{\mddefault}{\updefault}$\vdots$}}}}}
\put(2343,1575){\makebox(0,0)[lb]{\smash{{{\SetFigFont{7}{8.4}{\rmdefault}{\mddefault}{\updefault}$[a_{21},\dots,a_{2j_2},a]b_2$}}}}}
\put(768,450){\makebox(0,0)[lb]{\smash{{{\SetFigFont{7}{8.4}{\rmdefault}{\mddefault}{\updefault}$[a_{n1},\dots,a_{nj_n},a]b_n$}}}}}
\put(768,2325){\makebox(0,0)[lb]{\smash{{{\SetFigFont{7}{8.4}{\rmdefault}{\mddefault}{\updefault}$[a_{11},\dots,a_{1j_1},a]b_1$}}}}}
\put(2643,975){\makebox(0,0)[lb]{\smash{{{\SetFigFont{7}{8.4}{\rmdefault}{\mddefault}{\updefault}$\vdots$}}}}}
\put(2643,0){\makebox(0,0)[lb]{\smash{{{\SetFigFont{7}{8.4}{\rmdefault}{\mddefault}{\updefault}$\vdots$}}}}}
\end{picture}
}
  \end{center}
  \caption{Edges after the composition performed by HC.}
  \label{fig-HC-after}
\end{figure}
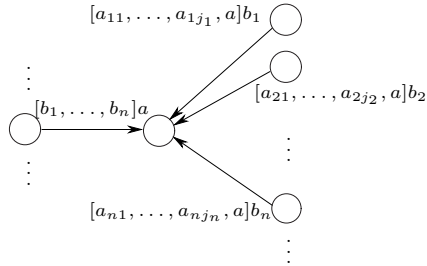

\begin{proof} Consider an edge $\al$ having positive weight $a$ as in the
Definition \ref{HCdef}, and suppose we stay in the second
case of the Definition, all the composable edges with
non positive weights coincident with $\al$ are the $\ba_i$'s with
weights $[a_{i1}, \dots ,a_{ij_i}]b_i$ for $i=1, \dots ,n$ as in
Figure \ref{fig-HC}.

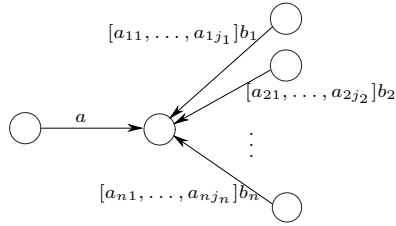
\begin{figure}[ht]
  \begin{center}
    \leavevmode
    \setlength{\unitlength}{0.00055000in}
\begingroup\makeatletter\ifx\SetFigFont\undefined%
\gdef\SetFigFont#1#2#3#4#5{%
  \reset@font\fontsize{#1}{#2pt}%
  \fontfamily{#3}\fontseries{#4}\fontshape{#5}%
  \selectfont}%
\fi\endgroup%
{\renewcommand{\dashlinestretch}{30}
\begin{picture}(4099,2122)(0,-10)
\put(2643,1502){\ellipse{300}{300}}
\put(2643,1950){\ellipse{300}{300}}
\put(1436,892){\ellipse{302}{302}}
\put(2650,149){\ellipse{284}{284}}
\put(158,906){\ellipse{300}{300}}
\blacken\path(1693.730,787.335)(1578.000,831.000)(1659.566,738.011)(1647.054,783.171)(1693.730,787.335)
\path(1578,831)(2515,182)
\blacken\path(1675.290,1036.544)(1585.000,952.000)(1704.453,984.108)(1658.410,992.828)(1675.290,1036.544)
\path(1585,952)(2493,1457)
\blacken\path(1612.595,1109.257)(1543.000,1007.000)(1652.528,1064.476)(1605.693,1062.907)(1612.595,1109.257)
\path(1543,1007)(2513,1872)
\path(318,907)(1268,907)
\blacken\path(1148.000,877.000)(1268.000,907.000)(1148.000,937.000)(1184.000,907.000)(1148.000,877.000)
\put(2253,1206){\makebox(0,0)[lb]{\smash{{{\SetFigFont{7}{8.4}{\rmdefault}{\mddefault}{\updefault}$[a_{21},\dots,a_{2j_2}]b_2$}}}}}
\put(2283,636){\makebox(0,0)[lb]{\smash{{{\SetFigFont{7}{8.4}{\rmdefault}{\mddefault}{\updefault}$\vdots$}}}}}
\put(940,1768){\makebox(0,0)[lb]{\smash{{{\SetFigFont{7}{8.4}{\rmdefault}{\mddefault}{\updefault}$[a_{11},\dots,a_{1j_1}]b_1$}}}}}
\put(858,240){\makebox(0,0)[lb]{\smash{{{\SetFigFont{7}{8.4}{\rmdefault}{\mddefault}{\updefault}$[a_{n1},\dots,a_{nj_n}]b_n$}}}}}
\put(632,967){\makebox(0,0)[lb]{\smash{{{\SetFigFont{7}{8.4}{\rmdefault}{\mddefault}{\updefault}$a$}}}}}
\end{picture}
}
  \end{center}
  \caption{Edges in a semi-full node.}
  \label{fig-HC}
\end{figure}

If we apply a step of the HC strategy by performing a DVR step between
$\al$ and $\ba_i$ for $1\le i \le n$, we obtain
$$\dvrstep(\al,\ba_i)=(\beta'_i,\al'_i)$$ where the weight of $\al$ is
$\fil{b_1,\dots,b_n}a$ and the weight of $\ba_i$ is $[a_{i1} \dots
a_{ij_i},a]b_i$ and the two new edges $\beta'$ and $\al'$ have a
positive weight, see Figure \ref{fig-HC-after}.

Therefore, now the set of the coincident filtered edges has been
enlarged with $\al$, but $\al$ is no more composable with the
$\ba_i$'s because of its filter and all the generated edges $\al_i$'s
and $\ba_i$'s have positive weights by definition of DVR. As a
consequence, all the coincident filtered edges (including $\al$) are
not composable with each other. Thus the obtained directed virtual net
is semifull.

If we stay in the first case of the definition \ref{HCdef}, performing
the composition of $\beta$ with all its non-positive coincident
counter-edges we obtain the same configuration as in the previous
case, moreover we compose $\alpha$ with (the residual of) $\beta$ and
so all the non-positive edges incident in the node are not further
composable. Note that the set of non-positive edges composable with
$\beta$ can possibly be empty, in this case HC just composes $\alpha$
and $\beta$.

\end{proof}

We recall that the translation presented in Section \ref{trans}
associates with any $\lambda$-term a full directed virtual net (see
also \cite{DanosRegnier93,AbadiGonthierLevy92}).  As full nets are
particular instances of semifull ones, HC actually represents a
reduction mechanism for $\lambda$-calculus.

Beyond the exploitation of parallelism, another interesting property
of HC is that we can separate the edges ending on a node in two
distinguished sets.
In other words, the strategy associates a mark with each edge: 
\emph{incoming} or \emph{combusted}.  
When created, edges are marked as incoming. One
step of reduction consists of picking an incoming edge $\alpha$ and
performing all the compositions with coincident combusted edges. Then
$\alpha$ is marked as combusted.

Note that an edge may be marked as combusted even when it has a
positive weight, namely if all the combusted edges coincident with
$\alpha$ are not composable with $\alpha$, with the particular case
where the set of combusted edges coincident with $\alpha$ is empty as
in case 1 of Definition \ref{HCdef}. On the other hand, at any step any
incoming edge has a positive weight.  As an edge is marked combusted
only after having been (successfully or not) composed with every
coincident combusted edges, one easily sees that two combusted edges
are never composable.  Thus this suggests that we can organize the
computation in such a way that the only meaning associated with
filters is about the belonging of an edge to the first or to the
second set (thus, like in the combustion strategy, filters can be
actually discarded). We have embedded this simplification among others
in the parallel implementation we present in the next section.

\section{The Implementation}\label{algorithm}

This section is devoted to the description of the implementation of
PELCR and is organized as follows. We first provide the outline of
data structures we have used and the high level description of the
parallel program. Then we enter details on any aspect and/or any
optimization characterizing the implementation. Actually, the material
presented in this section describes the implementation independently
of the specific language used to develop it (the {\tt C} language for
our case).

\subsection{Data Structures and Code Organization}

Each processor $i$ of the architecture hosting PELCR runs a process
$P_i$ which is an instance of the executable code associated with the
parallel program. We assume there is a master process, that for the
sake of clarity will be identified as $P_0$. All the other processes
will be referred to as slave processes. Processes communicate
exclusively by exchanging messages and the communication channels
among processes are assumed to be FIFO (this is not a limitation as
the most widely used message passing layers, such as PVM or MPI,
actually provide the FIFO property to communication channels). We call
\emph{pending} message any message already stored in the communication
channel, which has not yet been received by the recipient process.

We associate with each node $v$ an identifier, namely $id(v)$. Each
edge $e=(v_1,v_2)$ is therefore associated with the pair of node
identifiers $(id(v_1),id(v_2))$ thus the weighted edge is represented
by the triple $(id(v_1),id(v_2), W(e))$. 

As discussed in Section \ref{optrules}, by Property
\ref{orthogonalsystems} any edge $e$ incident on a node $v$ can be
seen as belonging to one of two distinct sets depending on which
between the two edges having $v$ as source originated $e$ through
composition. We call the two sets of edges as LEFT set and RIGHT set,
and we associate with each edge $e$ an additional information, namely
$Side(e)$, indicating whether $e$ belongs to the LEFT or the RIGHT
set. This information allows us to reduce the number of edge
compositions according to the HC strategy which must be performed
during the computation. Specifically, given two edges $e$ and $e'$
incident on a same node $v$, if the side of the two edges is the same,
no composition involving $e$ and $e'$ must be performed at all since
we a priori know that it will produce null result. On the other hand,
if $Side(e) \neq Side(e')$, composition must be performed to determine
the result, which can be either null or non-null.

\begin{figure}
\centerline{
\begin{picture}(0,0)%
\epsfig{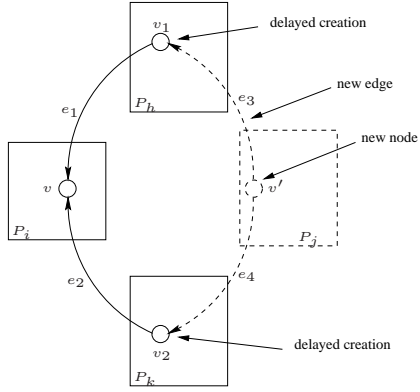}%
\end{picture}%
\setlength{\unitlength}{2013sp}%
\begingroup\makeatletter\ifx\SetFigFont\undefined%
\gdef\SetFigFont#1#2#3#4#5{%
  \reset@font\fontsize{#1}{#2pt}%
  \fontfamily{#3}\fontseries{#4}\fontshape{#5}%
  \selectfont}%
\fi\endgroup%
\begin{picture}(4362,4749)(64,-4048)
\put(1651,-3961){\makebox(0,0)[lb]{\smash{\SetFigFont{5}{6.0}{\rmdefault}{\mddefault}{\updefault}$P_k$}}}
\put(151,-2161){\makebox(0,0)[lb]{\smash{\SetFigFont{5}{6.0}{\rmdefault}{\mddefault}{\updefault}$P_i$}}}
\put(1651,-586){\makebox(0,0)[lb]{\smash{\SetFigFont{5}{6.0}{\rmdefault}{\mddefault}{\updefault}$P_h$}}}
\put(751,-661){\makebox(0,0)[lb]{\smash{\SetFigFont{5}{6.0}{\rmdefault}{\mddefault}{\updefault}$e_1$}}}
\put(2926,-511){\makebox(0,0)[lb]{\smash{\SetFigFont{5}{6.0}{\rmdefault}{\mddefault}{\updefault}$e_3$}}}
\put(3301,-1636){\makebox(0,0)[lb]{\smash{\SetFigFont{5}{6.0}{\rmdefault}{\mddefault}{\updefault}$v'$}}}
\put(526,-1636){\makebox(0,0)[lb]{\smash{\SetFigFont{5}{6.0}{\rmdefault}{\mddefault}{\updefault}$v$}}}
\put(1876,389){\makebox(0,0)[lb]{\smash{\SetFigFont{5}{6.0}{\rmdefault}{\mddefault}{\updefault}$v_1$}}}
\put(1876,-3661){\makebox(0,0)[lb]{\smash{\SetFigFont{5}{6.0}{\rmdefault}{\mddefault}{\updefault}$v_2$}}}
\put(3676,-2236){\makebox(0,0)[lb]{\smash{\SetFigFont{5}{6.0}{\rmdefault}{\mddefault}{\updefault}$P_j$}}}
\put(826,-2761){\makebox(0,0)[lb]{\smash{\SetFigFont{5}{6.0}{\rmdefault}{\mddefault}{\updefault}$e_2$}}}
\put(2926,-2686){\makebox(0,0)[lb]{\smash{\SetFigFont{5}{6.0}{\rmdefault}{\mddefault}{\updefault}$e_4$}}}
\end{picture}
}
  \caption{Creation of a new node.}
  \label{dvrp}
\end{figure}

In the general case, each process $P_i$ hosts only a subset of the
nodes of the graph. Therefore, given an edge $e=(v_1,v_2)$, there is
the possibility that $v_1$ and $v_2$ are hosted by distinct processes.
In Figure \ref{dvrp} we show an example of this. The interesting point
in the example is that when process $P_i$ performs the composition
between the edges $e_1$ and $e_2$ incident on node $v$ according to
HC, then a new node, namely $v'$ is originated together with two new
edges, namely $e_3$ and $e_4$ incident on nodes $v_1$ and $v_2$
respectively. The new node $v'$ can be hosted by any process, and
process $P_i$ is the one which establishes where $v'$ must be actually
located; in our example, $P_i$ selects $P_j$. We will come back to the
selection issue in Section \ref{balancing} when describing the load
balancing module that establishes how new nodes must be distributed
among processes. Note that, in case one of the newly produced edges
should have weight one, the {\em optimization of one rule} described
in Section \ref{optrules} (see Figure \ref{oneopt}), allows
avoiding the real creation of that edge. Also, the only edge really
created has as source a node already within the directed virtual
net, thus no new node needs to be created and addressed to some process.

In our implementation $id(v')$ is a triple $[t,P_i,P_j]$ where $P_i$
is the process that created the node $v'$, $P_j$ is the process
hosting that node and $t$ is a time-stamp value assigned by $P_i$.  The
time-stamp is managed by $P_i$ as follows: it is initialized to zero
and anytime $P_i$ originates a new node, it is increased by one.

When the new node $v'$ is originated by $P_i$, the creation must be
notified to $P_j$. Furthermore, both $P_k$ and $P_h$ must be notified
of the new edges $e_3$ and $e_4$ incident, respectively, on $v_1$ and
$v_2$.  In our implementation we use message exchange only for the
notification of new edges, while we avoid to explicitly notify the
creation of the new node $v'$ to $P_j$.  Process $P_j$ will actually
create the node $v'$ upon the receipt of the first message notifying a
new edge incident on $v'$. We will refer to this type of node creation
as {\em delayed creation}.  It allows us to reduce the amount of
notification messages exchanged among processes.

Applying the delayed creation technique to the example in Figure
\ref{dvrp} means that node $v_1$ is created by $P_k$ only upon the
receipt of the message carrying the information of the edge $e_3$
incident on $v_1$ (recall that this message is sent by $P_i$).
Similarly, $P_h$ will create $v_2$ only upon the receipt of the
notification message for the edge $e_4$ (also this message is sent by
$P_i$).

By previous considerations we get that any message exchanged between
two processes carries the information of a new edge.  Specifically, a
message carrying the information associated with the edge $e(v_1,v_2)$
has a payload consisting of the tuple
$[[t,P_i,P_j],[t',P_l,P_m],W(e),Side(e)]$ where $id(v_1)=[t,P_i,P_j]$,
$id(v_2)=[t',P_l,P_m]$, $W(e)$ is the weight of $e$ and $Side(e)$ is
the edge side.

$P_i$ keeps track of information related to local nodes in a list
$nodes_i$. Any element in $nodes_i$ has a compound structure.  In the
remainder of the article we identify the structure in $nodes_i$
associated with a node $v$ as $nodes_i(v)$. As relevant field of the
structure $nodes_i(v)$ we have a list, namely $nodes_i(v).combusted$,
containing the edges incident on the node $v$ which have already been
composed (i.e. the combusted edges of the HC strategy). The list
$nodes_i(v).combusted$ is partitioned into two sub-lists, namely
$nodes_i(v).combusted.LEFT$ and $nodes_i(v).combusted.RIGHT$,
containing edges having $Side()$ equal to LEFT and RIGHT respectively.

A buffer $incoming_i$ associated with $P_i$ is used to store received
messages. For what we have explained above, any message stored in
$incoming_i$ carries information related to a new edge which must be
added to the virtual net and composed with already combusted edges, if
any, incident on the same node.  Such an edge is actually an incoming
edge of the HC strategy. Therefore, the buffer $incoming_i$ represents
a kind of {\em work list} for process $P_i$, as, according to HC, any
incoming edge associated with a message stored in $incoming_i$
requires $P_i$ to compose it with all the already combusted edges
incident on the same node. Performing such a composition represents
the work associated with the message carrying the edge.

For each process $P_i$, except the master process $P_0$, both
$incoming_i$ and $nodes_i$ are initially empty, meaning that initially
there is no node of the directed net managed by $P_i$, nor there are
incoming edges for it. Instead, $P_0$ is such that its list $nodes_0$
is empty but its buffer $incoming_0$ contains a set of messages, one
for each initial edge of the virtual net (recall that the initial
edges are all incoming). Note that this does not mean $P_0$ is a
bottleneck for the parallel execution since the load balancing
mechanism we have implemented (see Section \ref{balancing}) promptly
distributes new edges produced in the early phase of the execution
among all the processes.

In Figure \ref{algo} we show the high level structure of the algorithm 
implemented by the software modules we have developed.
Before entering the pseudo code description, we recall that the HC strategy
is such that, any incoming edge of which process $P_i$ becomes aware
by extracting the corresponding message from $incoming_i$, must be
immediately composed with the preexisting edges incident on the same
node, without additional delay. Furthermore, given a message $m$
carrying the information of a new edge $e=(v_1,v_2)$, we denote as
$m.target$ the node identified by the information $id(v_2)$ carried by
$m$ (recall that $id(v_2)$ is the previously described triple) and as
$m.source$ the node identified by the information $id(v_1)$ carried on
the same message. $e.target$ and $e.source$ have similar meaning when
referring to an edge $e$. Also, we denote as $e_m$ the edge carried by
$m$.

The procedure $initialize()$ sets the initial values for all the data
structures. The procedure $empty()$ checks whether the buffer storing
received messages is empty. In the positive case, process $P_i$ has no
work to be performed, thus it invokes the procedure
$check\_termination()$ to check if the computation is actually ended,
i.e. no message will arrive (in Section \ref{termination} we will
provide details on how the detection of the termination is
implemented).  In the negative case, it extracts a message from the
buffer $incoming_i$ and performs the composition of the corresponding
incoming edge.

\begin{figure}[t]
\begin{scriptsize}
\hrulefill
\begin{tabbing} 
----\=----\=----\=----\=----\=----\=----\=\kill 
{\bf program} $P_i$;\\
1 \>$initialize()$;\\
2  \> {\bf while}  not $end\_computation$ {\bf do}\\
3 \> \> $\langle$collect all incoming messages and store them in $incoming_i\rangle$\\
4  \> \> {\bf while} not $empty(incoming_i)$ {\bf do}\\
5 \> \>\> $\langle$extract a message $m$ from $incoming_i$$\rangle$;\\
6 \>\> \> {\bf if} $m.target \in nodes_i$ ~~~~'node already in the local list' \\
7  \> \> \> {\bf then}\\
8 \>\>\> \> {\bf for} each edge $e \in nodes_i(m.target).combusted$ {\bf do} \\
9 \>\>\>\>\> {\bf if} $Side(e_m)\neq Side(e)$\\
10 \>\>\>\>\> {\bf then}\\
11\>\>\>\>\>\>$\langle$compose $e_m$ with $e$$\rangle$;\\
12 \> \> \> \> \> \> $\langle$select the destination process $P_j$ for hosting the node possibly\\
\>\>\>\>\>\>~~originated by the composition$\rangle$;\\
13  \>\> \> \> \>\>  $\langle$send the edges produced by the composition to $P_k$ and $P_h$ \\
\>\>\>\>\>\>~~hosting $m.source$ and $e.source$ respectively$\rangle$\\
14 \>\>\>\> {\bf endfor}\\
15 \> \> \> {\bf else} $\langle$add $m.target$ to $nodes_i\rangle$; ~~'delayed creation'\\
16 \>\>\>  $\langle$add  $e_m$ to $nodes_i(m.target).combusted.Side(e_m)\rangle$\\
17  \> \> {\bf endwhile};\\
18   \>\>    $\langle$$end\_computation = check\_termination()\rangle$;\\ 
19 \> {\bf endwhile}
\end{tabbing}
\hrulefill
\end{scriptsize}
\caption{Pseudo code for process $P_i$.}
\label{algo}
\end{figure}

By the test in line 11 we exploit information about $Side(e_m)$ to
avoid unnecessary edge compositions. The pseudo code structure also points
out that process $P_i$ checks for the presence of pending messages
only when $incoming_i$ is empty (i.e. when $P_i$ has no more work to
be performed unless new pending messages carry it).  This behavior
aims at reducing the communication overhead.  Specifically, a
procedure to check whether there are pending messages is realized
typically by using {\tt probe} functions supported by the used
communication layer. $P_i$ invokes the execution of a {\tt probe}
function to test if there is at least a pending message. If there is
at least one such message, then a {\tt recv} procedure is executed to
receive that message and store it into $incoming_i$. As pointed out in
other contexts \cite{DNRD96}, {\tt probe} functions may be expensive,
therefore, they should be executed only when a further delay could
actually produce negative effects on performance. In the general case,
delaying the {\tt probe} call until all the messages stored in
$incoming_i$ have been processed should not produce negative effects.
This is the reason why, in the general case, we suggest to perform the
{\tt probe} call only when $incoming_i$ becomes empty. However, we
noted that depending on the particular hardware/software architecture
and on the adopted message passing layer, excessive delays in receiving pending
messages could impact negatively on the performance of the
communication layer due to buffer saturation. This is the case we have
observed for our implementation based on MPI. For this reason we have
done a light modification to the general code structure in Figure
\ref{algo} in order to avoid excessively infrequent {\tt probe} calls
(and message receipts).

Beyond the overhead due to probe calls, another important issue is the
overhead related to send and receive operations. A solution to bound
this overhead will be discussed in the following subsection. Then we
will present the policy we have selected for balancing the load among
processes and other relevant aspects related to the implementation.

\subsection{Message Aggregation}
\label{aggr}

The cost of sending and receiving a physical message, paid part by the
sender and part by the receiver, can be divided into two components:
(i) an overhead that is independent of the message size, namely $oh$,
and (ii) a cost that varies with the size of the message, namely $s
\times oh_b$ where $s$ is the size (in bytes) of the message and
$oh_b$ is the send/receive time per byte. $oh$ typically includes the
context switch to the kernel, buffer reservation time, the time to
pack/unpack the message and, in case of distributed memory systems,
the time to setup the physical network path. Instead, $oh_b$ takes
into account any cost that scales with the size of the message.

$oh$ is usually higher than $oh_b$, as shown in \cite{XuHwang96} up to two
orders of magnitude, therefore it results usually more efficient to
deliver several information units (i.e. more than one application
message) with a single physical message, in such a way that a single
pair of send/receive operations is sufficient to download many data at
the recipient process. This  allows the reduction of the static
overhead $oh$ for each information unit, thus originating efficient
parallel executions, especially in the case of fine grain computations
like DVR.  As an example, if three application messages of size $s$
constitute the payload of a single physical message then the cost to
send and receive these application messages is reduced from $3oh + 3 s
\times oh_b$ to $oh + 3 s \times oh_b$.

We present below the optimization we have embedded in the
communication modules via the aggregation of application messages in a
single physical message. Each process $P_i$ collects application
messages destined to the same remote process $P_j$ into an aggregation
buffer $out\_buff_{i,j}$. Therefore, there is an aggregation buffer
associated with each remote process. Application messages are
aggregated and are infrequently sent via a single physical message.
The higher the number of application messages aggregated, the greater
the reduction of the static communication cost per application
message; we call this positive effect Aggregation Gain (AG). However,
the previous simple model for the communication cost ignores the
effects of delaying application messages on the recipient
process. More precisely, there exists the risk that the delay produces
idle times on the remote processes which have already ended their work
and are therefore waiting for messages carrying new work to be
performed; we call this negative effect Aggregation Loss
(AL). Previous observations outline that establishing a suited value
for the aggregation window (defined as the number of application
messages sent via the same physical message) is not a simple task.

In our implementation, the module controlling the aggregation keeps an
age estimate for each aggregation buffer $out\_buff_{i,j}$ by
periodically incrementing a local counter $c_{i,j}$. The value of
$c_{i,j}$ is initialized to zero and is set to zero each time the
application messages aggregated in the buffer are sent. At the end of
the composition phase of an incoming edge extracted from the local
work list $incoming_i$, $c_{i,j}$ is increased by one if at least one
message is currently stored in the aggregation buffer
$out\_buff_{i,j}$. Therefore, one tick of the age counter is equal to
the average combustion time of an incoming edge and the counter value
represents the age of the oldest message stored in the aggregation
buffer.

The simplest way to use previous counters is to send the aggregate
when the associated counter reaches a fixed value, referred to as
maximum age for the aggregate, or when the work list of the process is
empty.  In this case there is no need to delay the aggregate anymore
as the probability to put more application messages into it in short
time is quite small, so the delay will not increase AG and will
possibly produce an increase of AL. We will refer to this policy as
Fixed Age Based (FAB). Although this policy is simple to implement and
does not require any monitoring for the tuning of the maximum age over
which the aggregate must be sent, it may result ineffective whenever a
bad selection of the maximum age value is performed.

To overcome this problem we have implemented a Variable Age Based
(VAB) policy, which is an extension of FAB, having similarities with
an aggregation technique presented in \cite{CARW98} for communication
modules supporting fine grain parallel discrete event simulations. In
VAB, anytime the messages aggregated in $out\_buff_{i,j}$ are sent,
the message rate achieved by the aggregate is calculated. This rate is
used to determine what the maximum age for the next aggregate should
be. The dynamic change of the maximum age after which an aggregate
must be sent, allows the aggregation policy to adapt its behavior to
the behavior of the overlying application.

To implement VAB, $P_i$ is required to maintain an estimate
$est_{i,j}$ of the expected arrival rate in each aggregation buffer
$out\_buff_{i,j}$ (the higher such rate, the higher AG for that
buffer). This estimate can be computed by using statistics related to
a temporal window.  If the arrival rate for the current aggregate in
$out\_buff_{i,j}$ is higher than $est_{i,j}$ then the maximum age for
the next aggregate into that buffer is increased by one since the
application is likely to start a period of bursty exchange of
application messages from $P_i$ to $P_j$. Therefore a slight increase
in the maximum age is likely to relevantly increase AG.  If the
arrival rate falls below $est_{i,j}$, then the maximum age is decreased
by one (provided it is greater than one). An upper limit on the
maximum age can be imposed in order to avoid negative effects due to
AL (i.e. in order to avoid excessive delay for the delivery of the
aggregate at the recipient process).

\subsection{Load Balancing}\label{balancing}

Whenever the composition between two edges is performed by a process
$P_i$ then a new node is originated and $P_i$ must select a process
$P_j$ (possibly $j=i$) which will host the new node. In order to
provide good balance of the load we have implemented a selection
strategy for the destination process which uses approximated state
information related to the load condition on each process.

In our solution we identify the number of unprocessed application
messages $upm$ stored in the buffer $incoming_i$ as the state
information related to the load condition on $P_i$. $P_i$ keeps track
of the values of $upm$ related to itself and to the other processes
into a vector $UPM_i$ of size $n$ (where $n$ is the number of
processes).  $UPM_i[i]$ records the current value of the number of
unprocessed application messages of $P_i$. $UPM_i[j]$ records the
value of the number of unprocessed application messages of $P_j$ known
by $P_i$. These values are spread as follows.  Whenever $P_i$ sends a
physical message $M$ to $P_j$, the value of $UPM_i[i]$ is piggy-backed
on the message, denoted $M.UPM$ (\footnote{We use ``$M$'' to denote a
physical message in order to distinguish it from an application
message previously denoted as ``$m$''.}).  Whenever a physical message
$M$ sent by $P_j$ to $P_i$ is received from $P_i$, then $UPM_i[j]$ is
updated from $M.UPM$ (i.e. $UPM_i[j] \leftarrow M.UPM$). The
information on the load conditions kept by the $UPM$ vectors is
approximated for two reasons:
\begin{itemize}
\item there exists the possibility that when $P_i$ receives $M$ from 
$P_j$ the current value of $UPM_j[j]$ is different from $M.UPM$;

\item the current value of $UPM_i[i]$ is not 
an exact representation of the current load of $P_i$ as it does not count
application messages carried by pending physical messages; these
application messages represent work to be performed which has not yet
been incorporated into the buffer $incoming_i$.
\end{itemize}

We note, however, that obtaining more accurate state information on
the load condition of a process would require the exchange of
additional physical messages or, at worst, a synchronization among
processes which could produce unacceptable negative effects on the
performance.  Anyway, it is important to remark that the FIFO property
for the communication channels guarantees that each time a physical
message $M$ sent by $P_j$ is received from $P_i$, the piggy-backed
value $M.UPM$ refers to a more recent load condition as compared to
the one indicated by the current value of $UPM_i[j]$.

Based on the values stored in $UPM_i$, we have implemented a selection
policy for the destination process of a new node which is a modified
round-robin. It works as follows. $P_i$ keeps a counter $rr_i$
initialized to zero which is updated (module $n$) each time a new node
is produced by $P_i$. The current value of $rr_i$ is the identifier of
the process which should host the new node according to the
round-robin policy. $P_i$ actually selects $P_{{rr}_i}$ as destination
if $UPM_i[rr_i] < UPM_i[i]$; otherwise $P_i$ selects itself as
destination for the new node. In other words, each process distributes
the load in round-robin fashion unless, at the time the load
distribution decision must be taken, the local load is lower than that
of the remote process which should be selected.

\subsection{Termination Detection}\label{termination}

The implementation of the termination detection relies on the use of
additional control messages. Specifically, anytime a process distinct
from the master $P_0$ has no more work to be performed, it sends to
$P_0$ a message carrying information about both the number of
application messages received from other processes, which have already
been processed and the number of application messages produced for
other processes. Such a message will be referred to as {\tt status}
message. When the master $P_0$ detects that each process has already
elaborated all the application messages produced for it, $P_0$
discovers that the computation is over and notifies the termination to
the slave processes. This is done through the send of a {\tt
termination} message.  By looking at the structure of the code in
Figure \ref{algo}, it can be seen that process $P_i$ executes the
$check\_termination()$ procedure only when no work to be performed has
been detected (i.e. when $incoming_i$ is empty). This points out that
no synchronization is required (i.e. $P_i$ sends its {\tt status}
message without blocking to receive an acknowledgment; it will
possibly receive the {\tt termination} message during a future
execution of the $check\_termination()$ procedure)
(\footnote{Actually, to keep low the overhead due to {\tt status}
messages, $P_i$ sends one such message to the master only if its
status has changed since the last $\tt status$ message was sent.}). On
the other hand, the master $P_0$ checks for incoming {\tt status}
messages and possibly sends the {\tt termination} messages only when
it has no more work to be performed. Note that the consistency of the
information collected by $P_0$ through {\tt status} messages is
guaranteed by the FIFO property of communication channels.

\subsection{On-the-Fly Storage Recovery}

At the end of the computation we get that for all the nodes of the
final graph the incident edges are ghost.  However only some of those
nodes belong to the normal form of the reduction.  We recall that the
nodes of the initial graph belonging to the normal form are the nodes
initially having only ghost incident edges; this set of nodes will be
referred to as the {\em border}.  Starting from the border, we can
determine the whole normal form: it contains those nodes linked to the
border by a directed path.

We have embedded in the implementation a technique to discard
on-the-fly nodes that do not belong to the normal form.  We have taken
this design choice for keeping low memory usage with the twofold aim
of (i) increasing the efficiency of the underlying virtual memory
system, and (ii) allowing efficient management of the data structures
maintained at the application level. As respect to the latter issue,
anytime an unprocessed application message $m$ is extracted from the
buffer $incoming_i$ (see line 5 of the pseudo code in
Figure \ref{algo}), process $P_i$ must access information associated
with the node $m.target$, if it already exists. Such an information is
maintained in the structure $nodes_i(m.target)$ (see lines 6 on of the
pseudo code in Figure \ref{algo}). To retrieve the virtual memory
address for this structure we have used an hash table with chaining
for handling collisions, which keeps an active entry for each node in
the list $nodes_i$. Discarding nodes of valence zero that do not
belong to the normal form allows keeping low the number of entries of
the hash table, thus allowing efficient access to the table at
anytime.

The on-the-fly storage recovery technique we have implemented tracks
whenever a node becomes of valence zero and removes it if there is no
directed path towards nodes of the border. This is implemented through
a particular type of application messages we call {\tt EOT}
(End-of-Transmission) messages. Specifically, for each initial node
$v$ that does not belong to the border we insert an {\tt EOT} message.
If we ensure that {\tt EOT} messages are processed only after all the
messages carrying edges destined to $v$ have been already processed,
then we detect upon processing of the {\tt EOT} messages that no new
edge will have $v$ as its target. This means that node $v$, and
all the edges pointing to it, can be deleted. Before removing this
node, the {\tt EOT} message is propagated to the sources of the edges
pointing to $v$.  We note that, since each node is source of two
edges, we expect the arrival of two {\tt EOT} messages (one from both
sides) before handling the removal of the node. Therefore, if for a
node $v$ we have no {\tt EOT} message destined to it, or at most only
one of such messages, it means that $v$ has a directed path to the
border, thus it belongs to the normal form. In this case no removal
takes place.

Actually, guarantees that the {\tt EOT} messages will be processed
only after all the messages carrying incoming edges destined to the
same node have already been processed, is trivially achieved thanks to
the FIFO property of communication channels.

\section{Experimental Results}
\label{performance}

In this section we report experimental results demonstrating the
effectiveness of our implementation, and thus of both the HC strategy
underlying it and the combination of all the optimizations for the
run-time behavior we have presented and embedded within PELCR.  As
already pointed out, the implementation has been developed using the
{\tt C} language and MPI as the underlying message passing layer. A
major advantage of using such a standard interface for message passing
functionalities is that it makes the software highly portable. This
allowed us to test the implementation on a wide set of platforms such
as SMP machines with Linux, IBM mainframes like SuperPower 3 and
SuperPower 4 with AIX, shared memory Sun Ultra Sparc machines and
Alpha Digital Microchannel clusters.

In this section we report performance results obtained in the case of
an IBM server pSeries 690, with 32 Power4 CPUs - 1.3 GHz, 64 GB RAM,
running IBM AIX 5.2 ML1+ as the operating system. We have selected the
results obtained with this machine as representative especially
because of the larger number of available processors, as compared to
the other architectures. This allows us to better observe whether and
how the performance provided by PELCR scales while increasing the
computing power.

Before entering details related to the experimental results, we note
that there exists an approach to reduce the computation time in
optimal reduction systems based on an optimization known as {\em safe
operators} \cite{AspCh97}. It allows the merging of many control
operators in a compound one acting as the sequence, thus exhibiting
the ability to strongly decrease the number of interactions. Actually,
this approach has been the basis for the implementation of the so
called BOHM (Bologna Higher Order Machine), which is a sequential
machine for optimal reduction that has been demonstrated to provide
better performances compared to non-optimal interpreters such as CAML
and HASKELL (see \cite{AspertiGuerrini98}). Compared to this approach,
we tackle the issue of increasing the speed of the reduction in an
orthogonal way. Specifically, we do not use merging of operators to
reduce the number of interactions, instead we exploit computing
capabilities of multiple processors within the architecture to keep
low the reduction time via parallel computation of the reduction
itself. This approach can be applied to any computation issued from
the Geometry of Interaction. Note that we experimentally observe
speedup in parallel evaluation of terms typable in systems with
intrinsic complexity ELL (or LLL)
\cite{Girard95a}; these terms are evaluated in such a way
that the safe operators optimization is not needed. As a consequence, our
approach  is expected to speedup the execution
even for the cases in which safe operators cannot be effectively
employed, or even when optimal reduction does not affect 
the efficiency of computation like in non-higher order terms.

The results we report in this section refer to the following 
two different test cases:
\begin{itemize}
\item {\tt DD4}, which  corresponds to the
$\lambda$-term $(\delta)(\delta)\underline{4}$ where $\delta=\lambda x
(x)x$ represents the self application.  The normal form of this term
represents the Church's integer ${{(4^4)}^4}^4$.  
\item  {\tt EXP3}, which  corresponds to the ELL term 
$$\mathtt{Ite((Mult)2,1,Ite((Mult)2,1,Ite((Mult)2,1,4)))},$$
 whose normal form represents the
iterated exponential $2^{2^{2^4}}$. For a precise relation between 
this ELL term and the multiplicative linear logic proof net 
from which the dynamic graph to be executed is obtained, we refer the
reader to \cite{Ped96}. 
\end{itemize}

For both these two cases, the shared result of the HC strategy has a
number of nodes which is on the order of hundreds of thousands (for
{\tt DD4} this number even reaches about one million and half),
therefore they are large enough case studies to stress the behavior of
our implementation.

Before presenting the results, we provide details on the main
parameters we have measured in the experiments.

\subsection{Measured Parameters}
\label{parameters}

A measure of success of any parallel implementation is how
significantly it accelerates the computation. Typically the
acceleration is expressed by the so called {\em speedup}, evaluated as
the ratio between the sequential execution time on a single processor
and the parallel execution time on multiple processors. Actually, this
is a fundamental parameter to consider, not only because it expresses
the amount of increase in the execution speed while increasing the
power of the underlying computing system, but also because the speedup
curve provides indications on how the execution speed scales while
increasing the computing power. Linear speedup means that the
execution speed scales linearly vs the computing power. This is an
indication that the parallel implementation maintains the same
effectiveness independently of the number of used processors, thus the
implementation itself does not suffer, e.g., from excessive increase
in the communication overhead while increasing the number of processes
involved in the parallel execution. Actually, we also report the ratio
between the observed speedup and the ideal speedup that can be
achieved with a given degree of parallelism, i.e. with a given amount
of used processors. (We recall that the ideal speedup on $n$
processors is equal to $n$, which means we experience no overhead but
only gain by distributing the work to be performed on the $n$
processors.)  This parameter provides indications on the extent to
which the parallel implementation can be considered effective,
independently of the shape of the speedup curve.  Specifically, if we
have a linear speedup curve but a low ratio over the ideal speedup, it
means that the parallel implementation, although not particularly
suffering from increase in the overhead due to the parallelization
while increasing the amount of processors, is anyway ineffective,
e.g. due to inadequate structuring of the parallel algorithm it
implements.

Beyond speedup, another parameter we report is the {\em wall-clock
time} for the reduction.  This parameter expresses the real time cost
for a given reduction and also how it varies while increasing the
computing power of the underlying platform.  It is a fundamental
parameter to report since it provides indications on whether the
speedup curve has been evaluated over a representative interval for
what concerns the number of used processors. Specifically, if
wall-clock time values of few seconds or less are achieved while
increasing the number of processors, then it means that an additional
increase in the computing power does not make sense for this specific
reduction (this is because response time of few seconds or less is
typically considered satisfactory even for the case of an interactive
end-user, i.e. the case in which responsiveness is actually a critical
issue to address), hence the speedup has been evaluated over an
adequate interval for what concerns the degree of parallelism.

The wall-clock time and the speedup are parameters that express the
effectiveness of the parallel implementation when evaluated
globally. However, we are also interested in observing the effects of
specific optimizations we have proposed. As respect to this point, we
also report data that allow the evaluation of the benefits from the
VAB message aggregation technique discussed in Section \ref{aggr} and
the effectiveness of the load balancing policy presented in Section
\ref{balancing}. To evaluate how VAB impacts the communication cost while
increasing the amount of processors, we report the product between the
average number of application messages delivered through a single MPI
message, i.e. the average size of the aggregate which we will refer to
as AAS, and the number of used processors. This product is
representative of the system capacity to send application messages at
the time cost of sending a single MPI message. Specifically, when using
$n$ processors, the hosted processes can perform send operations of
MPI messages concurrently. Therefore, within the wall-clock time of a
single send operation, we are, on the average, able to send $n$ MPI
messages in parallel.  As a consequence, if the product between AAS
and the number of processors increases, we have that the time cost for
the send of each application message gets reduced, with consequent
reduction of the overall communication overhead on each
processor. (For completeness, we also report the plot for AAS, so as
to show its behavior while increasing the number of processors.) For
what concerns load distribution, we report plots related to the
variation, over time, of the amount of unprocessed application
messages, namely $upm$, stored in the $incoming$ buffer at different
processes (recall that $upm$ has been used in Section \ref{balancing} as the
information on current load on each processor to determine the
distribution of new nodes dynamically originated during the
computation). This parameter is representative of the effectiveness of
the load balancing strategy we have adopted since it provides
indications on whether the work list keeping track of the amount of
edges to be composed is approximatively the same on all processes at
any time during the execution.

\begin{figure}[t]
\begin{center}
\makebox(175,150){\scalebox{0.24}{\includegraphics{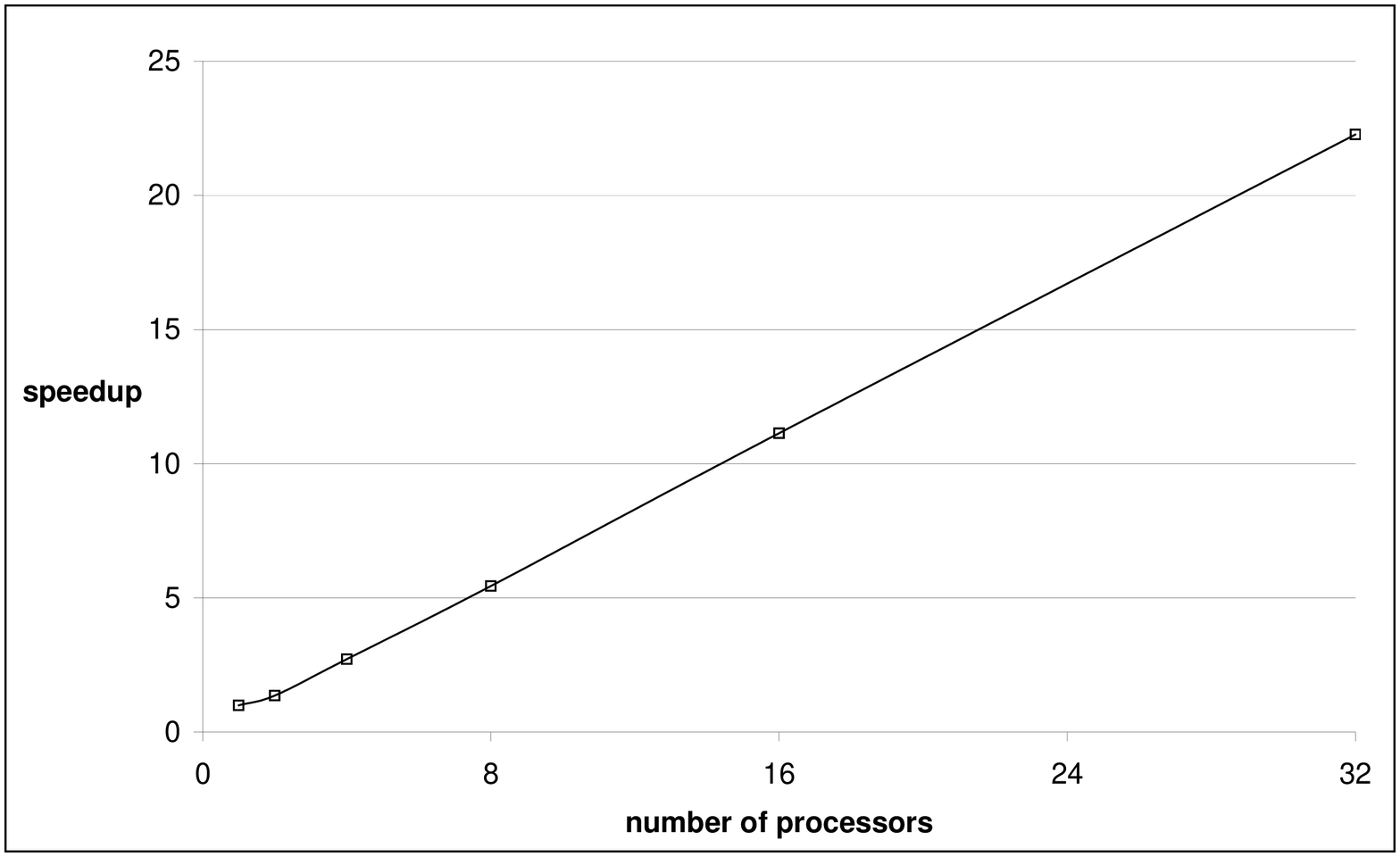}}}
\makebox(175,150){\scalebox{0.24}{\includegraphics{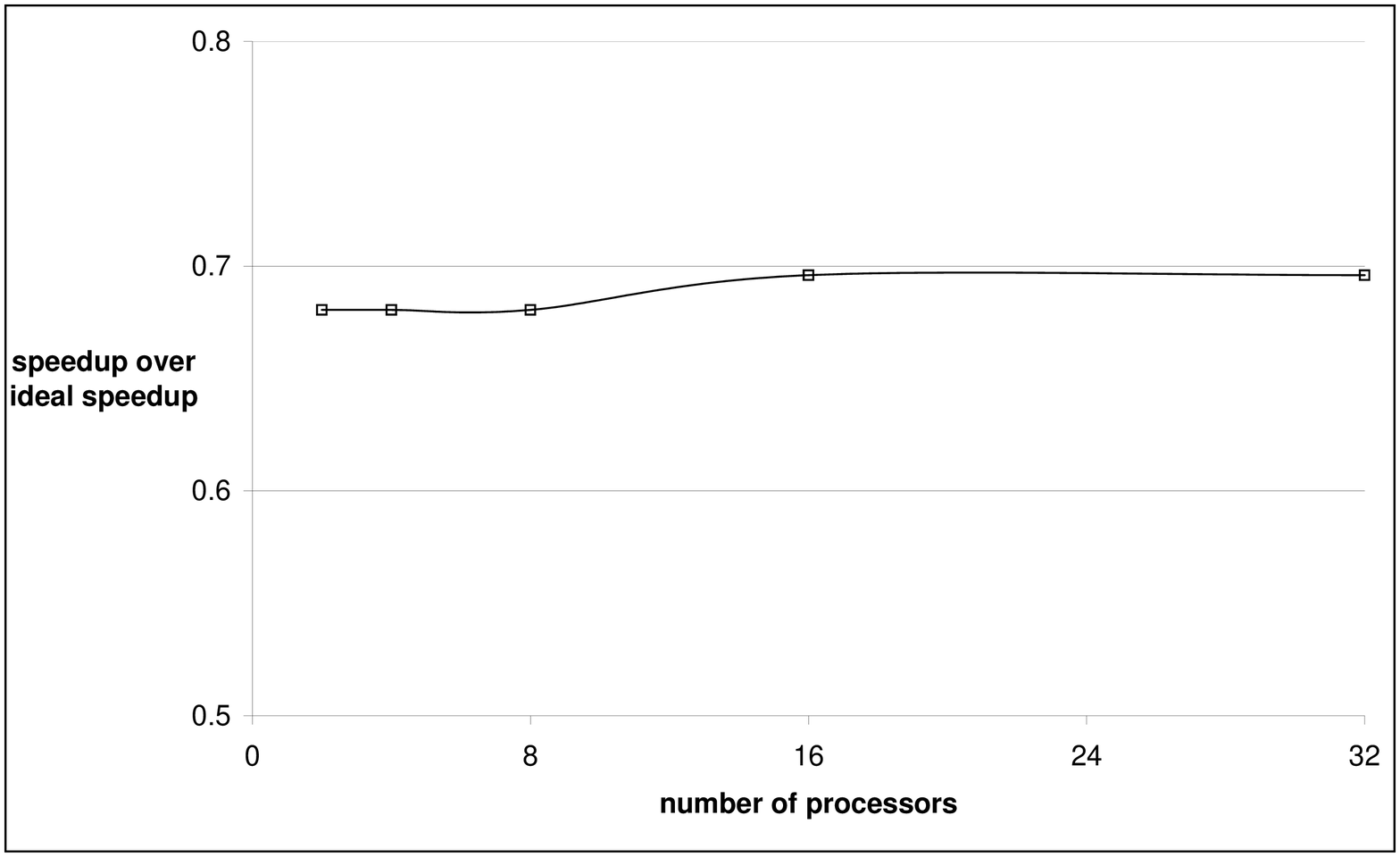}}}\\
\makebox(175,80){\scalebox{0.24}{\includegraphics{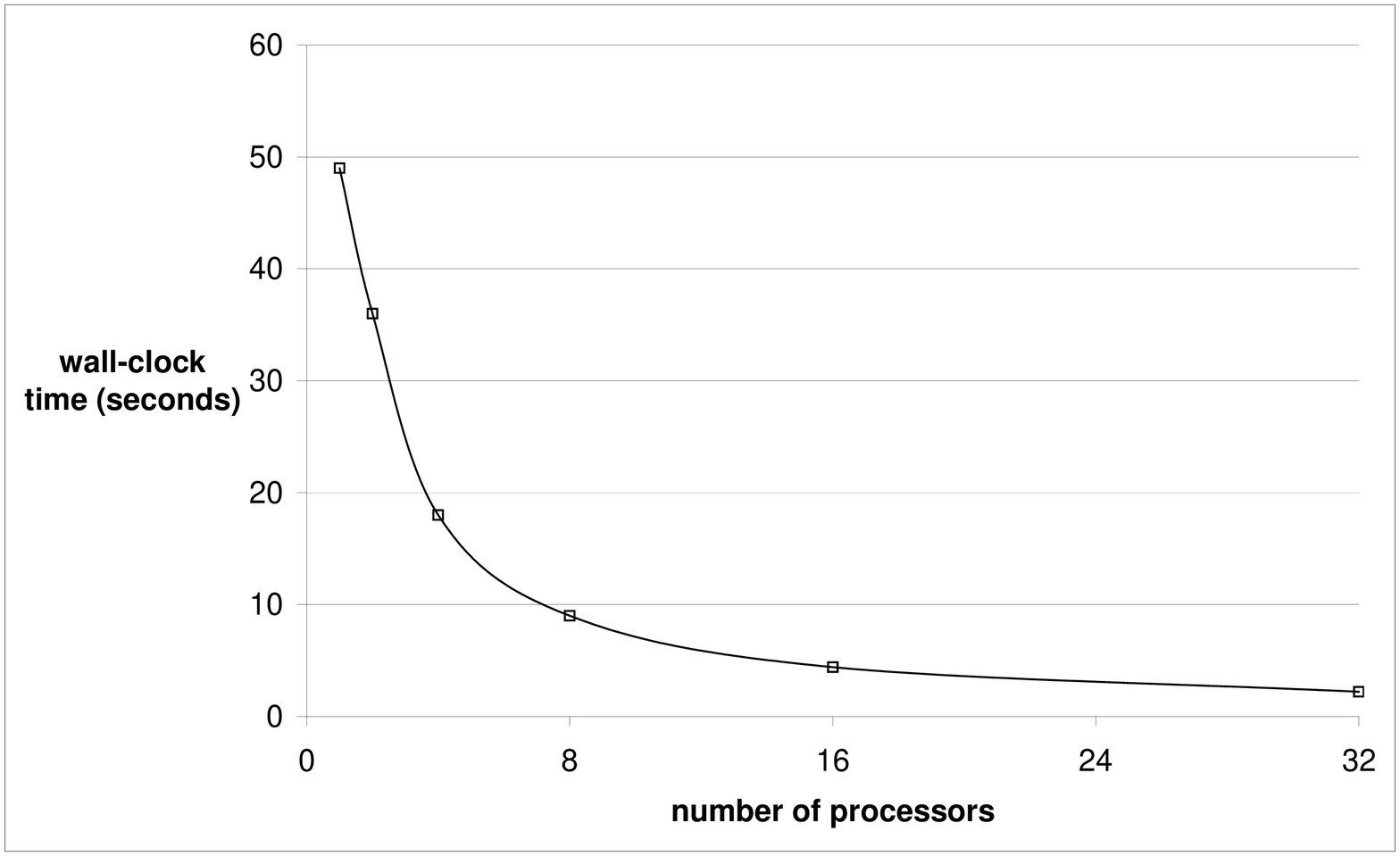}}}
\end{center}
\caption{Speedup and Wall-Clock Time results for {\tt DD4}.}
\label{dd4-speedup}
\end{figure}

\subsection{Results for {\tt DD4}}

\begin{figure}[t]
\begin{center}
\makebox(175,100){\scalebox{0.24}{\includegraphics{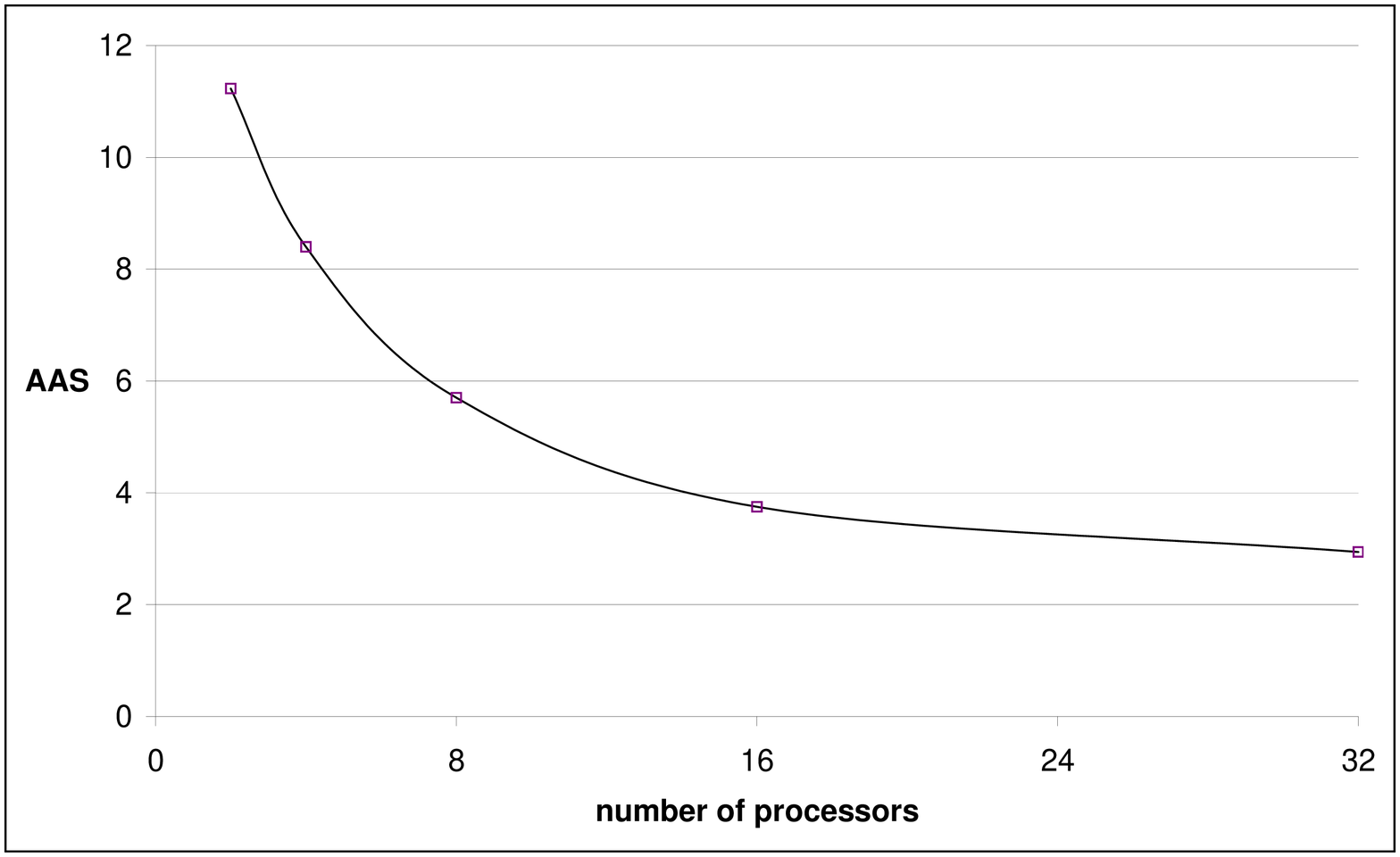}}}
\makebox(175,100){\scalebox{0.24}{\includegraphics{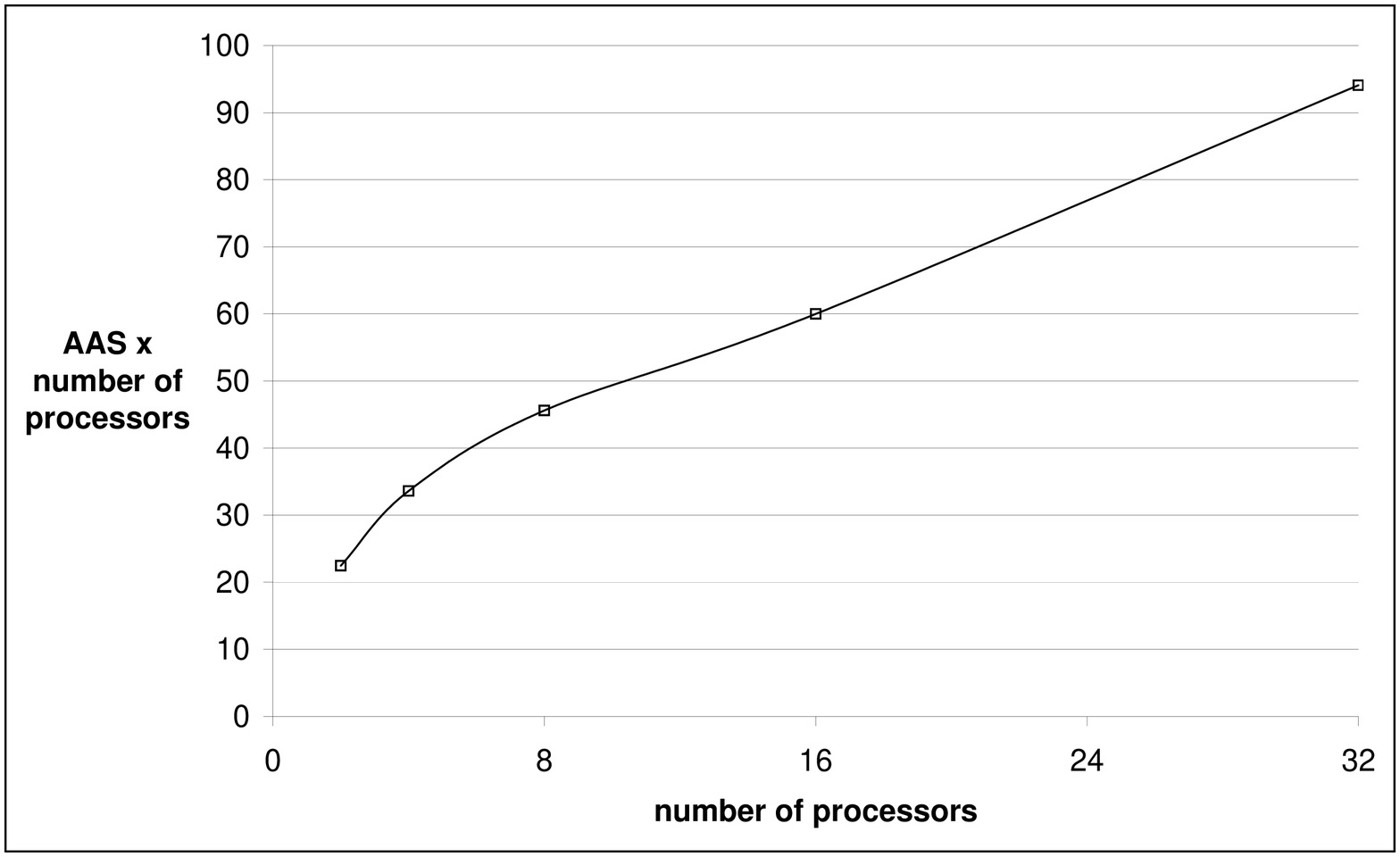}}}\\
\end{center}
\caption{VAB message aggregation results for {\tt DD4}.}
\label{dd4-aas}
\end{figure}

\begin{figure}[ht]
\begin{center}
\makebox(160,110)[lb]{\vspace{-15pt}
\hspace{-20pt}\scalebox{0.24}{\includegraphics{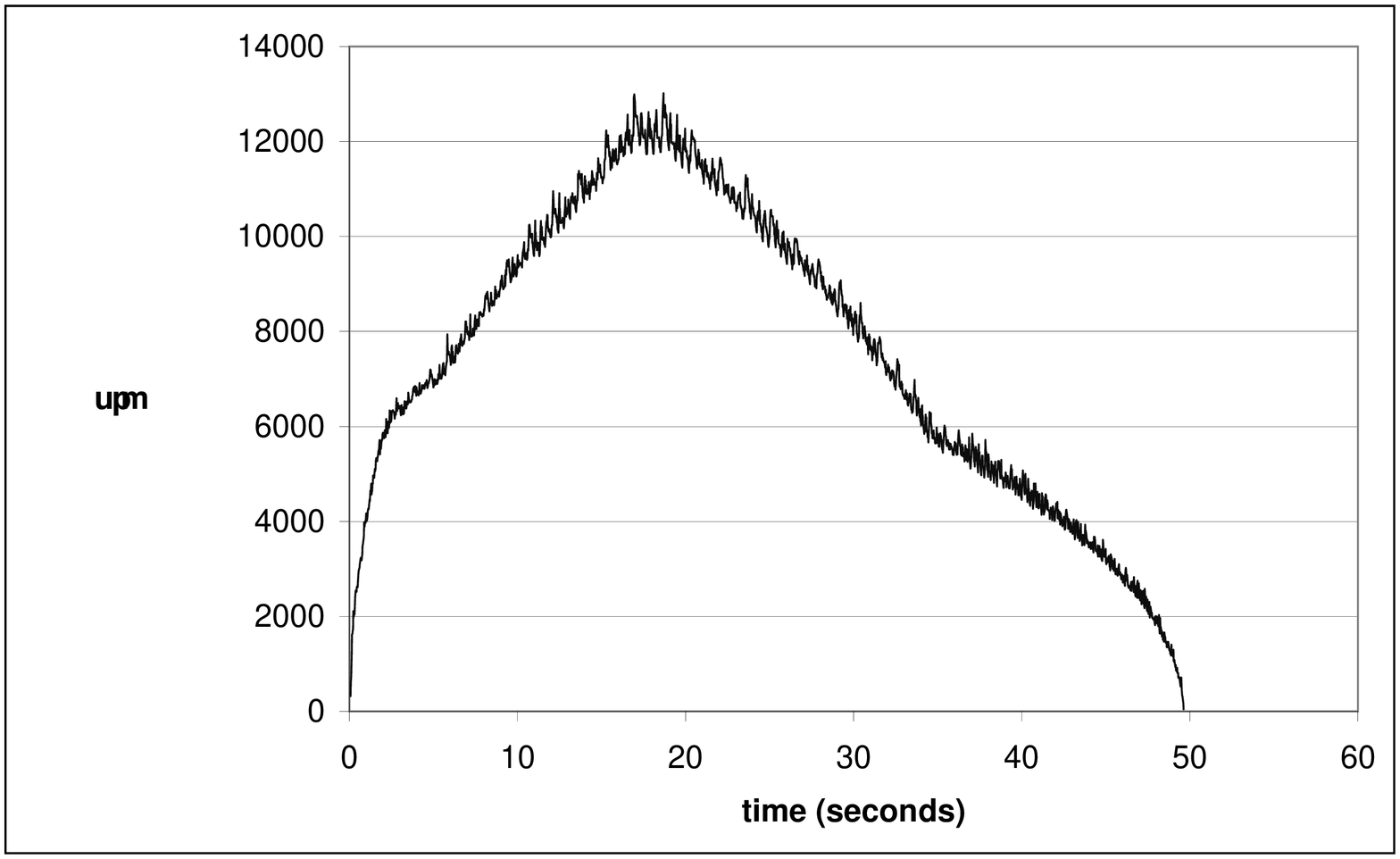}}}
\makebox(180,110)[lb]{\vspace{-96pt}

\hspace{-20pt}\resizebox{8cm}{8.8cm}{\includegraphics{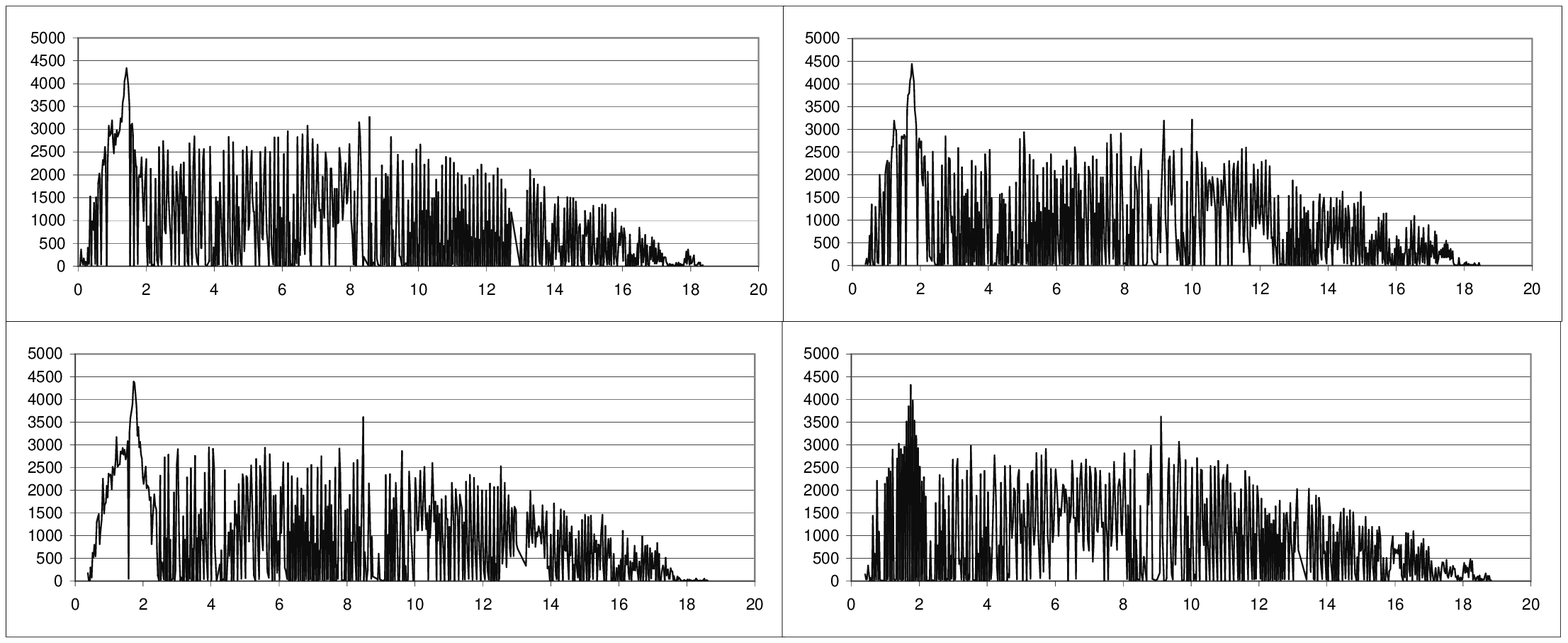}}}
\caption{Variation of $upm$ over time for {\tt DD4} 
(single processor case on the left - four processors case on the right).}
       \label{dd4-load}
\end{center}
\end{figure}

The experimental measures obtained for {\tt DD4} are reported in
Figures \ref{dd4-speedup}, \ref{dd4-aas} and \ref{dd4-load}. By the
plots in Figure \ref{dd4-speedup}, we observe that the speedup curve
remains linear over the whole interval for what concerns the number of
user processors (i.e. up to 32); also the speedup value is constantly
on the order of the 70\% of the ideal speedup. Combined toghether,
these two plots indicate that the parallel implementation is effective
for what concerns both the structuring of the parallel algorithm (this
provides the ability to reach high values with respect to the ideal
speedup) and the capability of remaining performance efficient while
increasing the amount of used processors. Also, the wall-clock time
curve, always reported in Figure \ref{dd4-speedup}, demonstrates that the
speedup plots provide reliable performance indications, in the sense
that speedup has been evaluated over an adequate interval for what
concerns the amount of used processors.  Specifically, with 32
processors, the wall-clock time for the computation gets on the order
of 2.2 seconds, which is not only a definitely reduced value as compared to the
sequential execution time (i.e. the execution time on a single
processor, namely about 50 seconds), but also represents a satisfactory 
response time for an interactive end-user.

The plots related to the behavior of the VAB aggregation technique in
Figure \ref{dd4-aas} and to the variation of $upm$ over time in Figure
\ref{dd4-load} additionally help understanding the reason why the
implementation remains effective while increasing the amount of used
processors. The AAS curve in Figure \ref{dd4-aas} shows that the
average amount of application messages aggregated within each MPI
message gets reduced from about 11.5 to about 3 while moving from the
single processor execution to the execution on 32 processors. This is
an expected behavior when thinking that a larger amount of used
processors means that each process $P_i$ involved in the parallel
execution needs to manage an increased amount of channels towards
other processes. As a consequence the application messages produced by
$P_i$ must be distributed over a larger amount of aggregation buffers
$out\_buff_{i,j}$, which means a reduced capacity to aggregate
messages within a given time unit in each single buffer. However,
observing the curve related to AAS multiplied by the number of used
processors, always in Figure \ref{dd4-aas}, we have a clear indication
that the system capacity to send application messages at a given cost
linearly increases with the number of processors, with a slope of
about 0.8 (recall the ideal case for the curve of AAS multiplied by
the amount of processors would be for slope equal to 1). Specifically,
when moving from $k$ processors to $2k$ processors, the system
increases its capacity of sending application messages at the same
time cost of about 1.6 times, which is a clear indication that VAB
allows the communication overhead to scale well vs the size of the
underlying computing system. For what concerns the variation of $upm$
over time, in Figure
\ref{dd4-load} we report both the case of single processor execution
and the case of execution on four processors. By the plots we observe
that the load of unprocessed application messages, stored by each
process $P_i$ within the $incoming_i$ buffer, is well distributed on
each of the four processors during the whole execution period, thus
supporting the claim of the effectiveness of the load balancing
mechanism described in Section
\ref{balancing}.

\subsection{Results for {\tt EXP3}}

\begin{figure}[t]
\begin{center}
\makebox(175,150){\scalebox{0.24}{\includegraphics{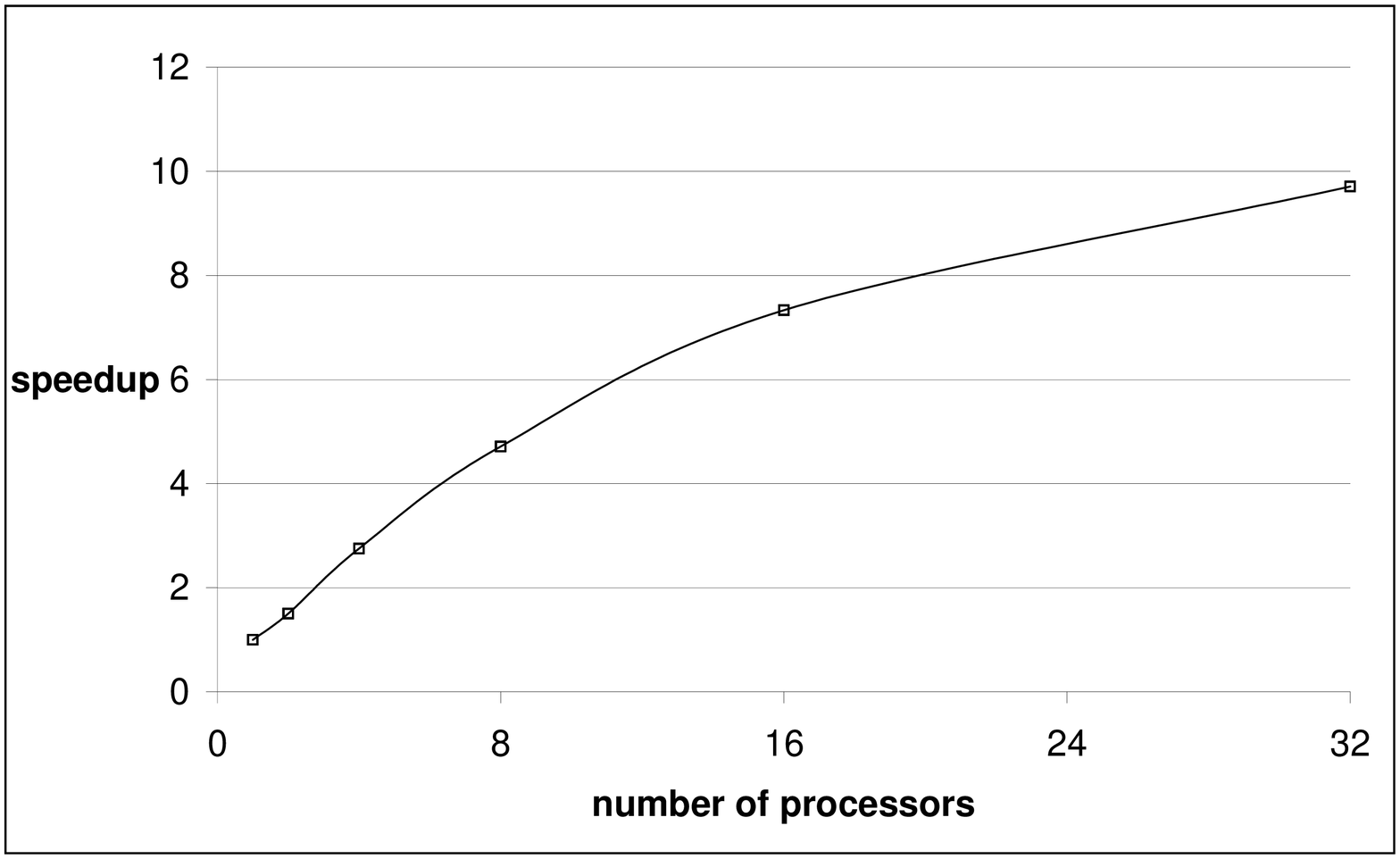}}}
\makebox(175,150){\scalebox{0.24}{\includegraphics{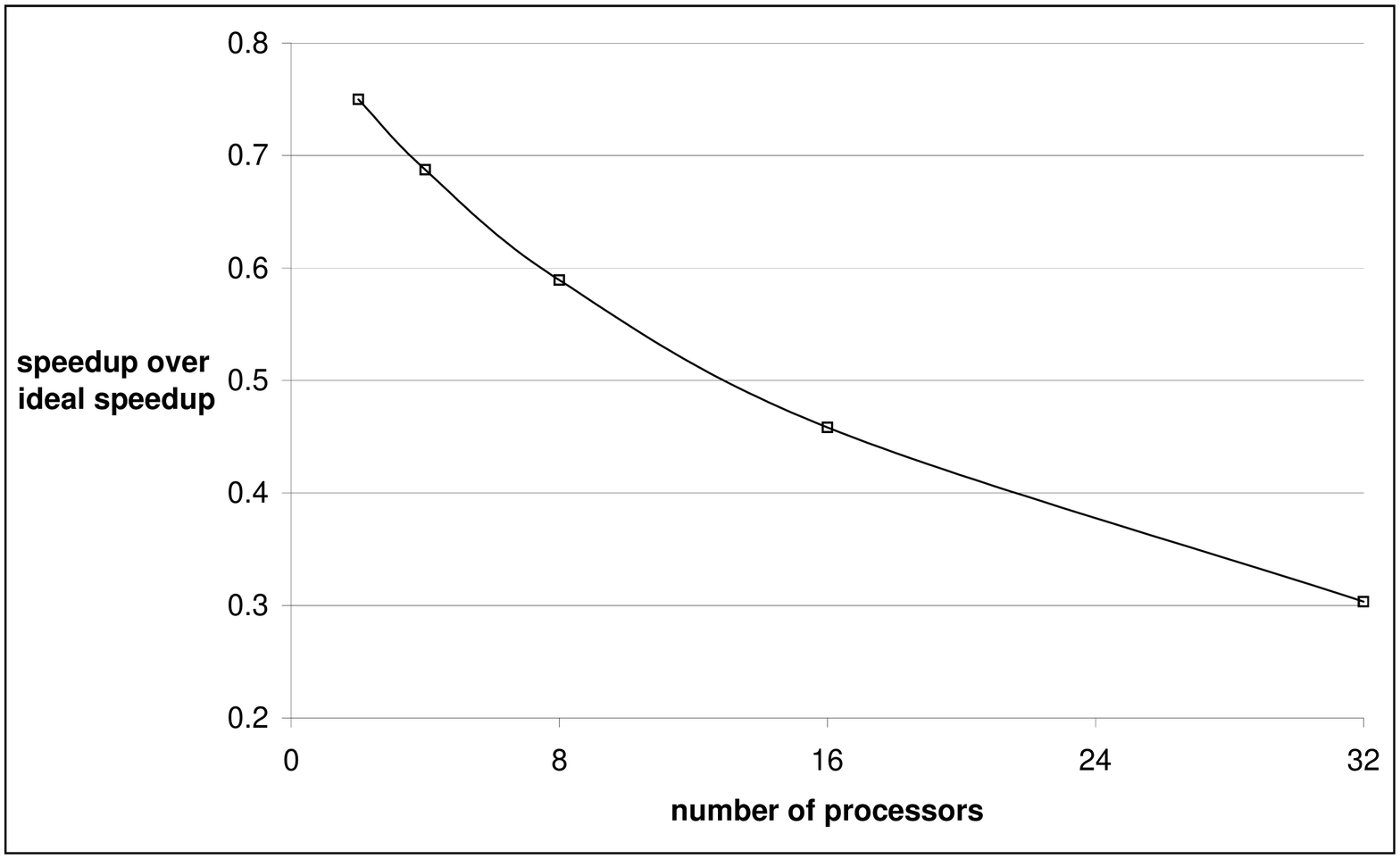}}}\\
\makebox(175,80){\scalebox{0.24}{\includegraphics{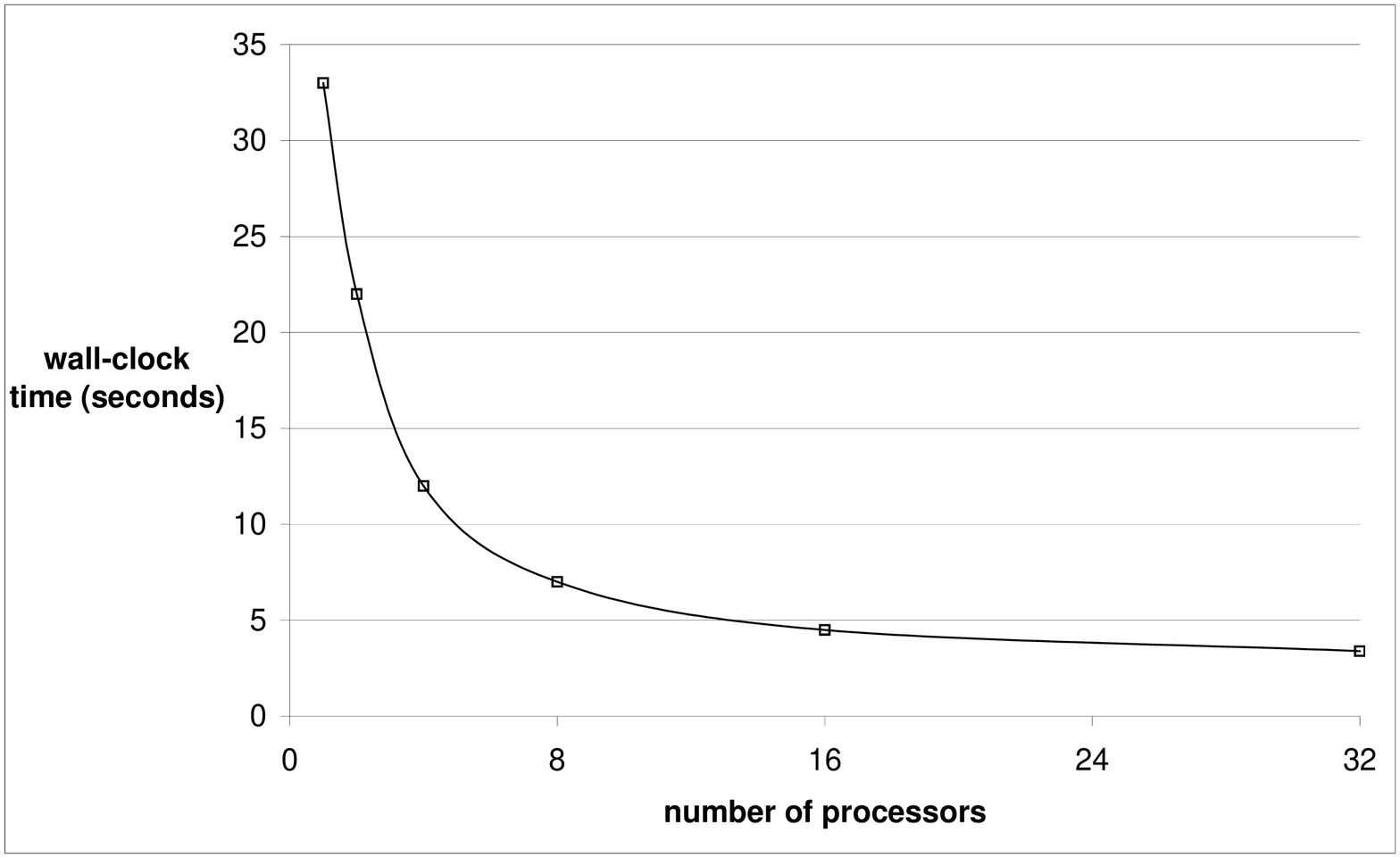}}}
\end{center}
\caption{Speedup and Wall-Clock Time results for {\tt EXP3}.}
\label{exp3-speedup}
\end{figure}

\begin{figure}[t]
\begin{center}
\makebox(175,100){\scalebox{0.24}{\includegraphics{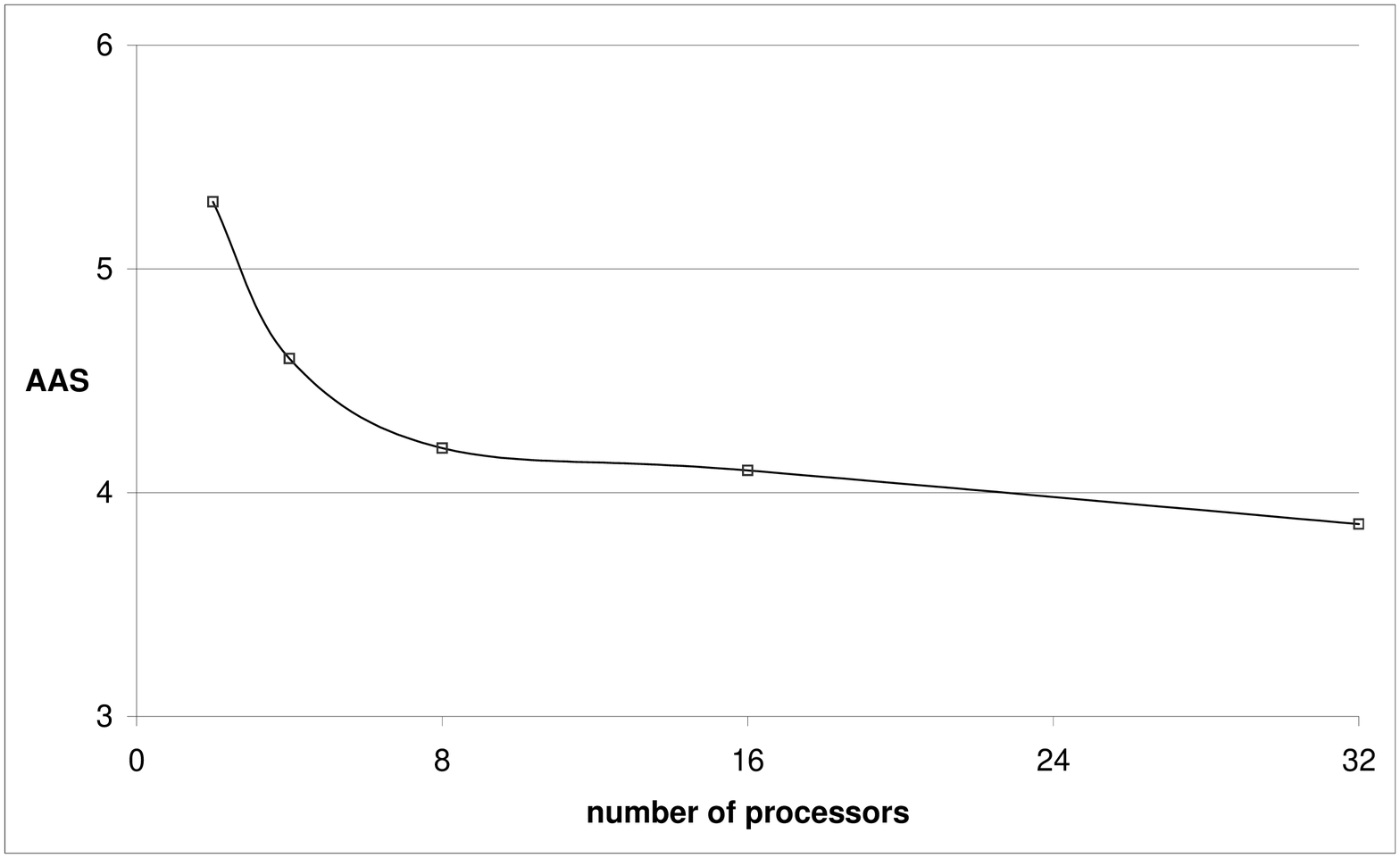}}}
\makebox(175,100){\scalebox{0.24}{\includegraphics{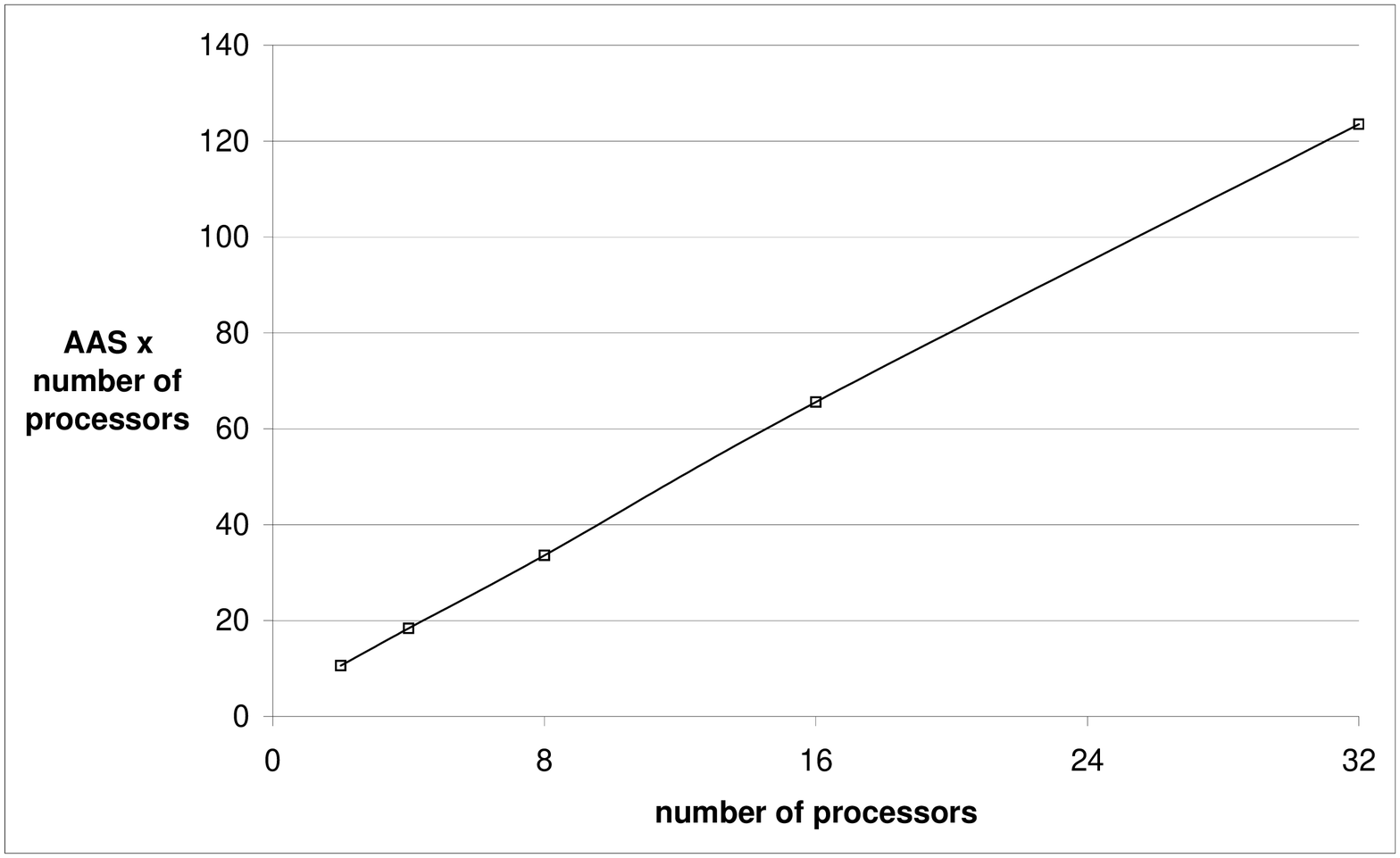}}}\\
\end{center}
\caption{VAB message aggregation results for {\tt EXP3}.}
\label{exp3-aas}
\end{figure}

\begin{figure}[ht]
\begin{center}
\makebox(160,110)[lb]{\vspace{-15pt}

\hspace{-10pt}\scalebox{0.20}{\includegraphics{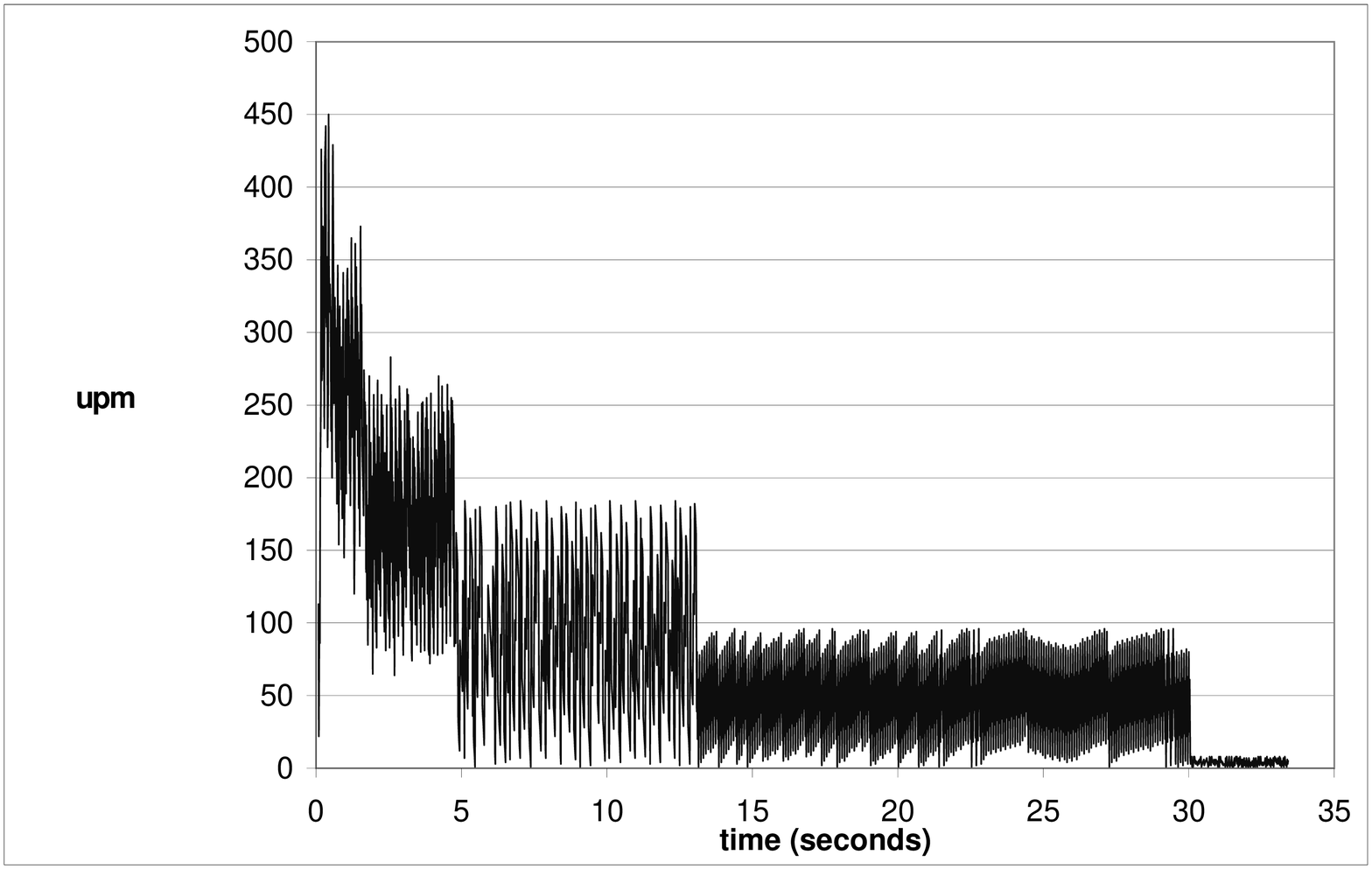}}}
\makebox(180,110)[lb]{\vspace{-417pt}

\hspace{-50pt}\resizebox{11cm}{22cm}{\includegraphics{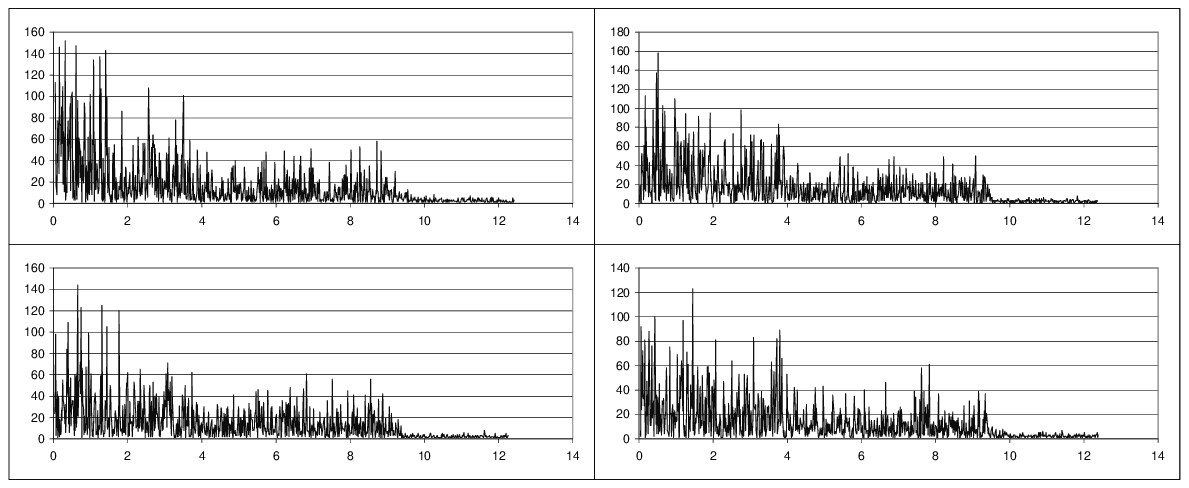}}}
\caption{Variation of $upm$ over time for {\tt EXP3} 
(single processor case on the left - four processors case on the right).}
        \label{exp3-load}
\end{center}
\end{figure}

The data obtained for the {\tt EXP3} application allow, in the light
of the already observed results for {\tt DD4}, the determination of
additional information on the run time behavior of our implementation,
also in terms of the effects of the particular application on the
achievable performance. The main difference with respect to the case
of {\tt DD4}, is in that this time the speedup curve does not remain
linear vs the number of processors. Specifically, the plots in Figure
\ref{exp3-speedup} show that the speedup asymptotically tends to a
constant value (on the order of 12), with a consequent decrease of the
ratio over the ideal speedup. By the data related to the effects of
VAB and to the variation of $upm$ over time, it can be deduced that
the cause for such a behavior is not due to ineffectiveness of the
parallel implementation (e.g. in terms of increase of the
communication overhead while increasing the amount of processors or
load imbalance). Specifically, the curve in Figure
\ref{exp3-aas} related to AAS multiplied by the number of used processors
clearly shows that also in this case the implementation is able to
carefully control the communication overhead while the size of the
underlying computing system gets increased. More precisely, the
linearity of such a curve, with slope on the order of about 1,
i.e. the ideal slope value, provides indication that the
implementation is able to control the communication overhead even in a
more effective way that what done for the case of {\tt DD4} in the
previous section (recall the slope for the same curve for {\tt DD4}
was 0.8, thus lower than what we observe in this case). Also, the
plots for $upm$ in Figure \ref{exp3-load} show that load remains
balanced while moving from the single processor case to the execution
on multiple processors.

Actually, that type of behavior for the speedup curve is due to the
fact that {\tt EXP3} exhibits a final computation phase made of a very
limited amount (i.e. few units) of unprocessed application
messages. This can be clearly observed when looking at the plots in
Figure \ref{exp3-load} related to the variation of $upm$ over time. In
other words, during that final phase the computation becomes intrinsic
sequential (few unprocessed application messages produce, once
extracted from the $incoming$ buffer and composed through HC, few new
application messages carrying new edges for the virtual net). As a
consequence, during the final phase, parallelism cannot be exploited
for {\tt EXP3}, which is exactly the reason why speedup asymptotically
tends to a constant value. This type of problem has been already
observed by Mackie in \cite{Mackie97} for the case of parallelism in
the form of adequate assignment of the initial nodes of the
interaction net over the used processors (recall this solution is
based on a static analysis of the initial interaction nets and differs
form our proposal in that we dynamically control load distribution,
and other performance indexes at run time). Specifically, also for
Mackie's approach, the benefits from the exploitation of multiple
processors in the computing system are bounded by phases of sequential
computation, if any, intrinsic to the specific application.  However,
as an additional support to the fact that our implementation is
definitely able to exploit parallelism, whenever present within
specific execution phases, we note that in case the speedup results
were computed by excluding the final sequential phase (by the plots in
Figure \ref{exp3-load} such a phase lasts about three seconds, which
is the reason why the wall-clock time asymptotically tends to three
seconds while increasing the number of used processors), they would be
even better that those obtained for the case of {\tt DD4} in the
previous section. Specifically, speedup would be on the order of at
least 75\% of the ideal over the whole interval for the amount of used
processors.

\section{Summary}

The definition of a formal system in \cite{DanosRegnier93,DPR97} for
the computation of the execution formula \cite{Girard89a} of the
$\lambda$-calculus terms constitutes the original starting point for
the work we have presented in this article.  Such a formal system was
initially motivated as the mathematical settlement of an operational
semantics for $\lambda$-calculus and for other functional programming
languages.  In making the precise definition of this system,
properties of locality and asynchrony pointed out the potential for
distributing the execution of programs.

The main contribution of this work relies on showing how it is
possible to make a functional language, based on $\lambda$-calculus,
transparently executable on a parallel/distributed environment. This
result has been achieved by exploiting the decomposition of
beta-reduction into a set of more elementary execution steps, each one
independent of the others, which gives an execution model extremely
flexible and very prone to be supported by a parallel/distributed
environment. Specifically, we exploited the properties of locality and
asynchrony of directed virtual reduction, namely the formal system
providing the previously mentioned decomposition to develop the PELCR
software package. This package manages the distribution of
computational load due to the evaluation of a $\lambda$-term in a
totally transparent way, by dynamically controlling/tuning any
parameter potentially affecting the efficiency of the run-time
behavior.

Our presentation integrates a solid theoretical background with many
practical techniques coming from current parallel computing
developments and provides a full featuring facility for the
parallel/distributed execution of functional programs.


\end{document}